\documentclass[%
reprint,
superscriptaddress,
showpacs,
amsmath,amssymb,
prc,
floatfix, ]%
{revtex4-1}

\usepackage{color}

\usepackage{graphicx}
\usepackage{dcolumn}
\usepackage{bm}
\usepackage[dvipdfm,bookmarks=true,colorlinks,%
            citecolor=blue,linkcolor=blue,hypertex, %
            breaklinks=true]{hyperref}

\begin{document}


\title{Theoretical study of the synthesis of superheavy nuclei with $Z=$ 119 and 120 
       in heavy-ion reactions with trans-uranium targets}

\author{Nan Wang}
 \affiliation{College of Physics, Shenzhen University, Shenzhen 518060, China}
\author{En-Guang Zhao}
 \affiliation{State Key Laboratory of Theoretical Physics,
              Institute of Theoretical Physics, Chinese Academy of Sciences,
              Beijing 100190, China}
 \affiliation{Center of Theoretical Nuclear Physics, National Laboratory
              of Heavy Ion Accelerator, Lanzhou 730000, China}
 \affiliation{School of Physics, Peking University,
              Beijing 100871, China}
\author{Werner Scheid}%
 \affiliation{Institut f\"ur Theoretische Physik der Justus-Liebig-Universit\"at,
              D-35392 Giessen, Germany}
\author{Shan-Gui Zhou}
 \email{sgzhou@itp.ac.cn}
 \affiliation{State Key Laboratory of Theoretical Physics,
              Institute of Theoretical Physics, Chinese Academy of Sciences,
              Beijing 100190, China}
 \affiliation{Center of Theoretical Nuclear Physics, National Laboratory
              of Heavy Ion Accelerator, Lanzhou 730000, China}

\date{\today}

\begin{abstract}
By using a newly developed di-nuclear system model with a dynamical potential
energy surface---the DNS-DyPES model, hot fusion reactions for synthesizing superheavy
nuclei (SHN) with the charge number $Z$~=~112--120 are studied.
The calculated evaporation residue cross sections are in good agreement with available data.
In the reaction $^{50}$Ti+$^{249}$Bk $\rightarrow ^{\ 299-x}$119 $+\ x$n, the maximal 
evaporation residue (ER) cross section is found to be about 0.11~pb for the 4n-emission channel.
For projectile-target combinations producing SHN with $Z=120$, the ER cross section 
increases with the mass asymmetry in the incident channel increasing.
The maximal ER cross sections for $^{58}$Fe+$^{244}$Pu and $^{54}$Cr+$^{248}$Cm
are relatively small (less than 0.01~pb) and those for $^{50}$Ti+$^{249}$Cf
and $^{50}$Ti+$^{251}$Cf are about 0.05 and 0.25~pb, respectively.
\end{abstract}


\maketitle



In the last decades, a lot of experimental progresses have been made in
synthesizing superheavy elements (SHE).
Up to now, SHEs with charge number $Z \le 118$ have been produced via cold fusion
reactions with Pb or Bi as targets~\cite{Hofmann2000_RMP72-733, Morita2004_JPSJ73-2593}
and hot fusion reactions with $^{48}$Ca as
projectiles~\cite{Oganessian2007_JPG34-R165, Oganessian2010_PRL104-142502}.
There have been also some attempts to synthesize superheavy nuclei (SHN) with $Z > 118$.
For example, experiments with projectile-target combinations
$^{58}$Fe+$^{244}$Pu~\cite{Oganessian2009_PRC79-024603} and
$^{50}$Ti+$^{249}$Cf~\cite{Dullmann2011_TASCA11} have been performed
in order to produce the element 120 but no $\alpha$ decay chains consistent
with fusion-evaporation reaction products were observed.

The evaporation residue (ER) cross section $\sigma_\mathrm{ER}$ of fusion
reactions depends strongly on the projectile-target combination and
the incident energy.
The study of such dependences is interesting and useful particularly
when one tries to synthesize new SHEs with $Z > 118$
because $\sigma_\mathrm{ER}$ of reactions with these nuclei as evaporation
residues becomes tiny which makes the experiment much more difficult.
In recent years much effort has devoted to the investigation of the synthesis
mechanism of SHN with $Z > 118$.
Using a dinuclear system (DNS) model,
Feng \textit{et al.} calculated the cross sections of cold fusion reactions with
isotopes of elements 119 and 120 as evaporation residues and
the maximal $\sigma_\mathrm{ER}$ was predicted to be about 0.03~pb and 0.09~pb for
elements 119 and 120, respectively~\cite{Feng2007_PRC76-044606}.
Hot fusion reactions with $^{50}$Ti as the projectile were studied extensively
by using the DNS models~\cite{Nasirov2009_PRC79-024606, Adamian2009_EPJA41-235,
Gan2011_SciChinaPMA54S1-61, Nasirov2011_PRC84-044612},
the fusion by diffusion models~\cite{Liu2009_PRC80-054608,
Siwek-Wilczynska2010_IJMPE19-500, Liu2011_PRC83-044613, Liu2011_PRC84-031602R},
and some other models~\cite{Zagrebaev2008_PRC78-034610, Wang2011_PRC84-061601R}.
The optimal incident energy and the maximal $\sigma_\mathrm{ER}$ from different models
or with different parameters vary very much.
For example, $\sigma_\mathrm{ER}$ for the reaction
$^{50}$Ti+$^{249}$Cf $\rightarrow\ ^{296}$120 $+\ 3$n ranges from 
1.5 to 760~fb~\cite{Nasirov2011_PRC84-044612}.

In this work, we will study hot fusion reactions producing
SHN with $Z \ge 112$ and in particular the elements 119 and 120
using a newly developed DNS model with a dynamical potential energy surface
(DyPES)~\cite{Zhou20100416_Giessen, Wang2012_in-prep} (the
model is termed as the DNS-DyPES model).
The importance of dynamical deformations of fragments in dissipative heavy-ion
collisions has been long known~\cite{Riedel1979_ZPA290-47,*Riedel1979_ZPA290-385, Wolschin1979_PLB88-35}.
The dynamical deformations of the projectile and the target in the entrance channel
have been included in the DNS models~\cite{Li2003_EPL64-750, Li2006_JPG32-1143, 
Wang2008_PRC78-054607} and
they are crucial for calculating the local excitation energy of a DNS
during the process of nucleon transfers.
Quite recently, Huang \textit{et al.}~\cite{Huang2011_PRC84-064619} developed a new DNS model
which takes into account the dynamical deformations of each DNS.
In that model, a three-dimensional master equation is solved with three variables,
the deformations $\beta_i$ ($i=1$ and 2 for each nucleus in a DNS) and
the mass asymmetry $\eta$.
In order to i) take into account the influence of the dynamical deformations of nuclei
in each DNS and ii) make the calculation easier, we treat the dynamical deformations
in a more transparent and economic way in the DNS-DyPES model. 
The details of this model will be published elsewhere~\cite{Wang2012_in-prep} and 
here we only discuss briefly how to include the dynamical deformation.

As usual, the evaporation residue cross section in a heavy-ion fusion reaction
is calculated as the summation over all partial wave $J$,
\begin{eqnarray}
 \sigma_\mathrm{ER}(E_\mathrm{c.m.}) 
 & = &
 \sum_{J} \sigma_\mathrm{cap}(E_\mathrm{c.m.},J)
          P_\mathrm{CN}(E_\mathrm{c.m.},J)
 \nonumber \\ 
 &   & \mbox{} \ \ \ \ \times
          W_\mathrm{sur}(E_\mathrm{c.m.},J)
 ,
 \label{eq:sigma_ER}
\end{eqnarray}
with $E_\mathrm{c.m.}$ the incident energy in the center of mass frame.
In this work the capture cross section $\sigma_\mathrm{cap}$
is calculated with an empirical coupled channel
approach~\cite{Zagrebaev2001_PRC64-034606, Feng2006_NPA771-50}
and the survival probability $W_\mathrm{sur}$ is calculated using
a statistic model~\cite{Zubov2002_PRC65-024308, Xia2011_SciChinaPAM54S1-109}.
In the fusion process, an excited compound nucleus (CN) may be formed.
We calculate the formation probability of a CN $P_\mathrm{CN}$
based on the DNS concept.
The DNS concept was proposed by Volkov in order to describe
the deep inelastic transfer process~\cite{Volkov1978_PR44-93}.
This concept was later used to study the competition between complete fusion
and quasi-fission and to calculate the fusion probability in fusion
reactions~\cite{Antonenko1993_PLB319-425, Antonenko1995_PRC51-2635, Adamian1997_NPA627-361}.
Based on the DNS concept, several models have been developed for the study of
the synthesis mechanism of SHN (see~\cite{Zhao2008_IJMPE17-1937, Zubov2009_PPN40-847,
Li2010_NPA834-353c, Feng2011_NPR28-1} and references therein).

The basic idea of the DNS concept is that after the capture process,
a DNS $(A_\mathrm{P}, A_\mathrm{T})$ in the entrance channel is formed.
Then the DNS evolves via nucleon transfer along the mass
asymmetry coordinate $\eta$ instead of in the direction of the relative distance
between the projectile and the target $R$.
During the nucleon transfer process, any DNS $(A_1, A_2)$ with 
$A_1 = 0, 1, \cdots, A_\mathrm{P}+A_\mathrm{T}$ and 
$A_2 =  A_\mathrm{P}+A_\mathrm{T} - A_1$ may be formed.
The evolution of the distribution function of each DNS with time
can be described by a master equation~\cite{Li2003_EPL64-750, Li2006_JPG32-1143, Wang2008_PRC78-054607},
\begin{widetext}
\begin{eqnarray}
 \frac{dP(A_1,t)}{dt}
 & = &
 \sum_{A'_1}
  W_{A_1A'_1}(t) \left[ d_{A_1}(t) P(A'_1,t) - d_{A'_1}(t) P(A_1,t) \right]
 -
 \Lambda^\mathrm{qf}_{A_1}(t) P(A_1,t)
  .
 \label{eq:ME}
\end{eqnarray}
\end{widetext}
Since $A_1+A_2 = A_\mathrm{P} + A_\mathrm{T}$, only $A_1$ is explicitly included
in the above equation.
$d_{A_1}(t)$ is the microscopic dimension for a DNS $(A_1,A_2)$ with a local 
excitation energy $E^*_\mathrm{DNS}$ defined in Eq.~(\ref{eq:LocalEx}). 
$E^*_\mathrm{DNS}$ is shared by the two nuclei in this DNS.
For each nucleus, a valence space is opened due to the excitation and those nucleons in 
the states within the valence space are active for the transfer between the two nuclei. 
$d_{A_1}(t) = C^{N_1}_{m_1} C^{N_2}_{m_2}$ where $N_k$
is the number of valence states and $m_k$ is 
that of valence nucleons~\cite{Li2003_EPL64-750}. 
$\Lambda^\mathrm{qf}_{A_1}$ is the quasifission probability of the DNS $(A_1,A_2)$ 
and $W_{A_1A'_1}(t) = W_{A'_1A_1}(t)$ is the mean transition probability
between the DNS's $(A_1,A_2)$ and $(A'_1,A'_2)$.
For the details about how to solve the master equation, please refer to
Refs.~\cite{Li2003_EPL64-750, Li2006_JPG32-1143, Wang2008_PRC78-054607}.
Here we only focus on the local excitation energy and the dynamical deformations.

In Eq.~(\ref{eq:ME}), $d_{A_1}(t)$,  $\Lambda^\mathrm{qf}_{A_1}$,
and $W_{A_1A'_1}(t)$ all depend on the local excitation energy of the DNS,
\begin{eqnarray}
 E^*_\mathrm{DNS}(A_1,t)
 & = &
 E_\mathrm{total} - E^\mathrm{0}_\mathrm{DNS}(A_1,t) - E^\mathrm{rot}_\mathrm{DNS}(t)
 ,
 \label{eq:LocalEx}
\end{eqnarray}
with
\begin{eqnarray}
 E_\mathrm{total} & = & E_\mathrm{c.m.} + (M_\mathrm{T} + M_\mathrm{P}) c^2,
 \\
 E^\mathrm{0}_\mathrm{DNS}(A_1,t) & = & V_\mathrm{DNS}(A_1,t) + (M_1 + M_2) c^2,
 \label{eq:E_DNS}
 \\
 E^\mathrm{rot}_\mathrm{DNS}(t) & = & \frac{J(J+1)}{2{\cal J}_\mathrm{DNS}(A_1,t)}.
\end{eqnarray}
$V_\mathrm{DNS}(A_1,t) =  V_\mathrm{N}(A_1,t) + V_\mathrm{C}(A_1,t)$
and $V_\mathrm{N}(A_1,t)$ and $V_\mathrm{C}(A_1,t)$ are the nuclear 
and the Coulomb interactions between the two nuclei.
The potential energy in the mass asymmetry degree of freedom,
which is often called as the driving potential at $t=0$, is defined as
\begin{equation}
 V_\mathrm{PES}(A_1,t) \equiv
 V_\mathrm{DNS}(A_1,t) + (M_1 + M_2 - M_\mathrm{T} - M_\mathrm{P}) c^2
 \label{eq:PES}
.
\end{equation}
The interaction potential $V_\mathrm{DNS}(A_1,R,t)$ between the two nuclei in
a DNS depends on the distance between their centers $R$ and $V_\mathrm{DNS}(A_1,t)$
in Eqs.~(\ref{eq:E_DNS}) and (\ref{eq:PES}) takes the minimum value of the pocket 
with respect to $R$ in $V_\mathrm{DNS}(A_1,R,t)$, \textit{i.e.},
$V_\mathrm{DNS}(A_1,t) \equiv V_\mathrm{DNS}(A_1,R,t) | _{R=R_\mathrm{pocket}}$.

%
\begin{figure}
\begin{centering}
\includegraphics[width=0.85\columnwidth]{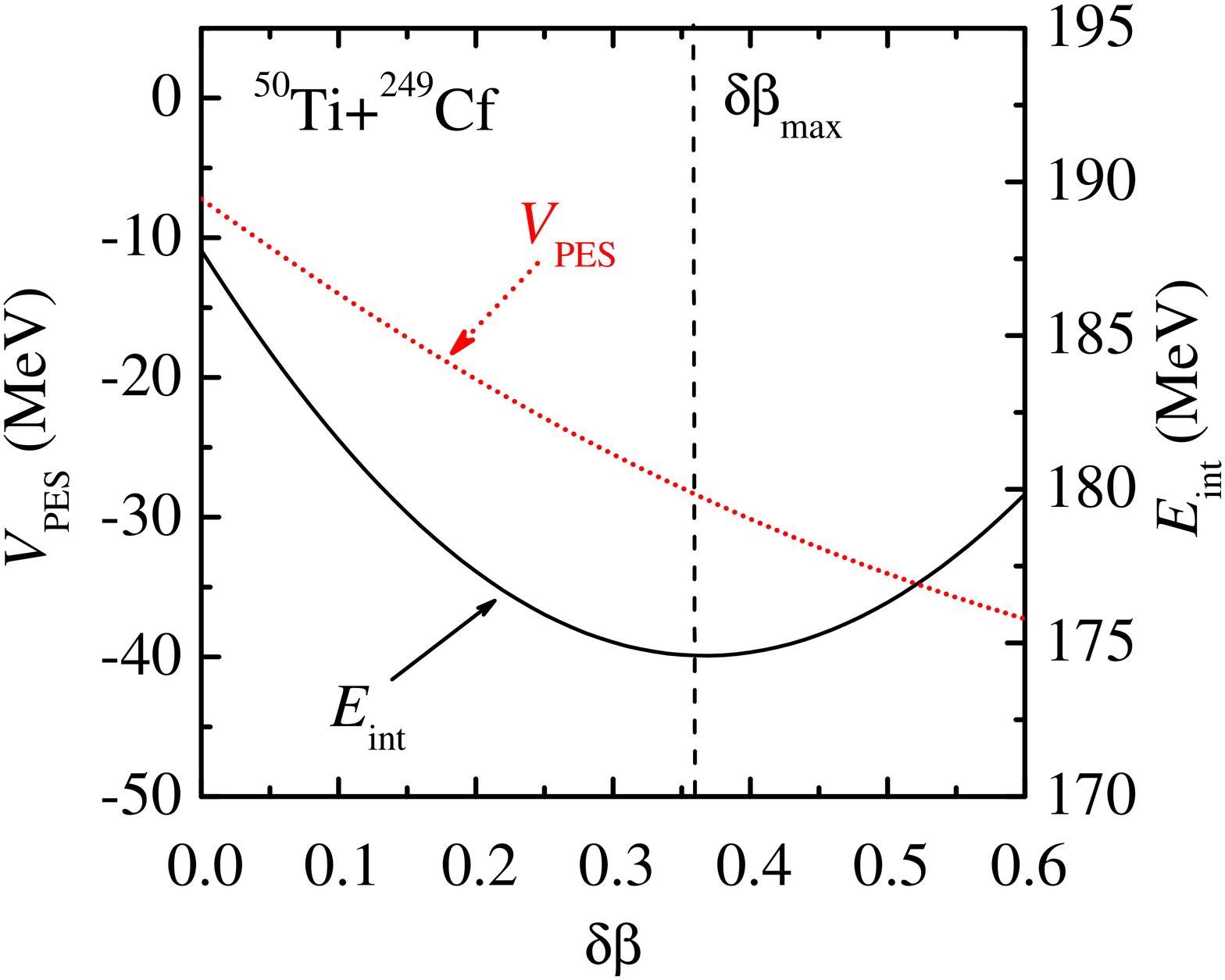}
\par\end{centering}
\caption{\label{fig:EV-db}(Color online)
The total ``intrinsic'' energy $E_\mathrm{int}$ (the black solid curve) and
the potential energy $V_\mathrm{PES}$ (the red dotted curve)
as a function of the dynamical deformation $\delta\beta$ for 
the projectile-target combination $^{50}$Ti + $^{249}$Cf.
The vertical dashed line shows the maximal dynamical deformation $\delta\beta_\mathrm{max}$.
}
\end{figure}

Due to the attractive nuclear force and the repulsive Coulomb force, both nuclei
in a DNS are distorted and dynamical deformations develop during the process of 
nuclear transfers~\cite{Riedel1979_ZPA290-47,*Riedel1979_ZPA290-385, Wolschin1979_PLB88-35}.
This results in the time-dependence of the potential energy surface (PES).
The nuclear interaction is calculated with a double-folding method~\cite{Adamian1996_IJMPE5-191} and
the Coulomb interaction from the Wong formula~\cite{Wong1973_PRL31-766}.
In this work we assume a tip-tip orientation of the two deformed nuclei and
that the dynamical deformations of the two nuclei satisfy
$\delta\beta^2_1 C_1/A_1 = \delta\beta^2_2 C_2/A_2$~\cite{Zagrebaev2001_PRC64-034606}
with the stiffness parameter $C_i$ ($i = 1$ and 2) calculated from a liquid drop 
model~\cite{Myers1966_NP81-1}.
We then define $\delta\beta \equiv (\delta\beta_1 + \delta\beta_2)/2 $ and
following Refs.~\cite{Riedel1979_ZPA290-47,*Riedel1979_ZPA290-385, Wolschin1979_PLB88-35} 
we assume that the dynamical deformation evolves in an overdamped motion,
\begin{equation}
 \delta \beta (t) = \delta\beta_\mathrm{max} \left( 1 - e^{-t/\tau_\mathrm{def}} \right)
 ,
\end{equation}
where the relaxation time $\tau_\mathrm{def} = 40 \times 10^{-22}$ s~\cite{Wolschin1979_PLB88-35} 
and the maximal dynamical deformation $\delta\beta_\mathrm{max}$ is determined by
minimizing the total ``intrinsic'' energy,
\begin{eqnarray}
 E_\mathrm{int}(A_1,\delta\beta)
 & = &
 V_\mathrm{N}(A_1;\beta_1,\beta_2) + V_\mathrm{C}(A_1;\beta_1,\beta_2) 
 \nonumber \\
 & + &
 \sum_{i=1,2} \frac{1}{2} C_i \delta\beta^2_i
 ,
\end{eqnarray}
where the quadrupole deformation $\beta_i = \beta^0_i + \delta\beta_i$ ($i=1$ and 2) with 
the static deformation parameters $\beta^0_i$ taken from Ref.~\cite{Moeller1995_ADNDT59-185}.
This is illustrated for $^{50}$Ti + $^{249}$Cf in Fig.~\ref{fig:EV-db} (the black solid curve) 
where one finds
that with $\delta\beta$ increasing, $E_\mathrm{int}$ decreases and takes the minimal value
at $\delta\beta \sim 0.36$.

\begin{figure}
\begin{centering}
\includegraphics[width=0.70\columnwidth]{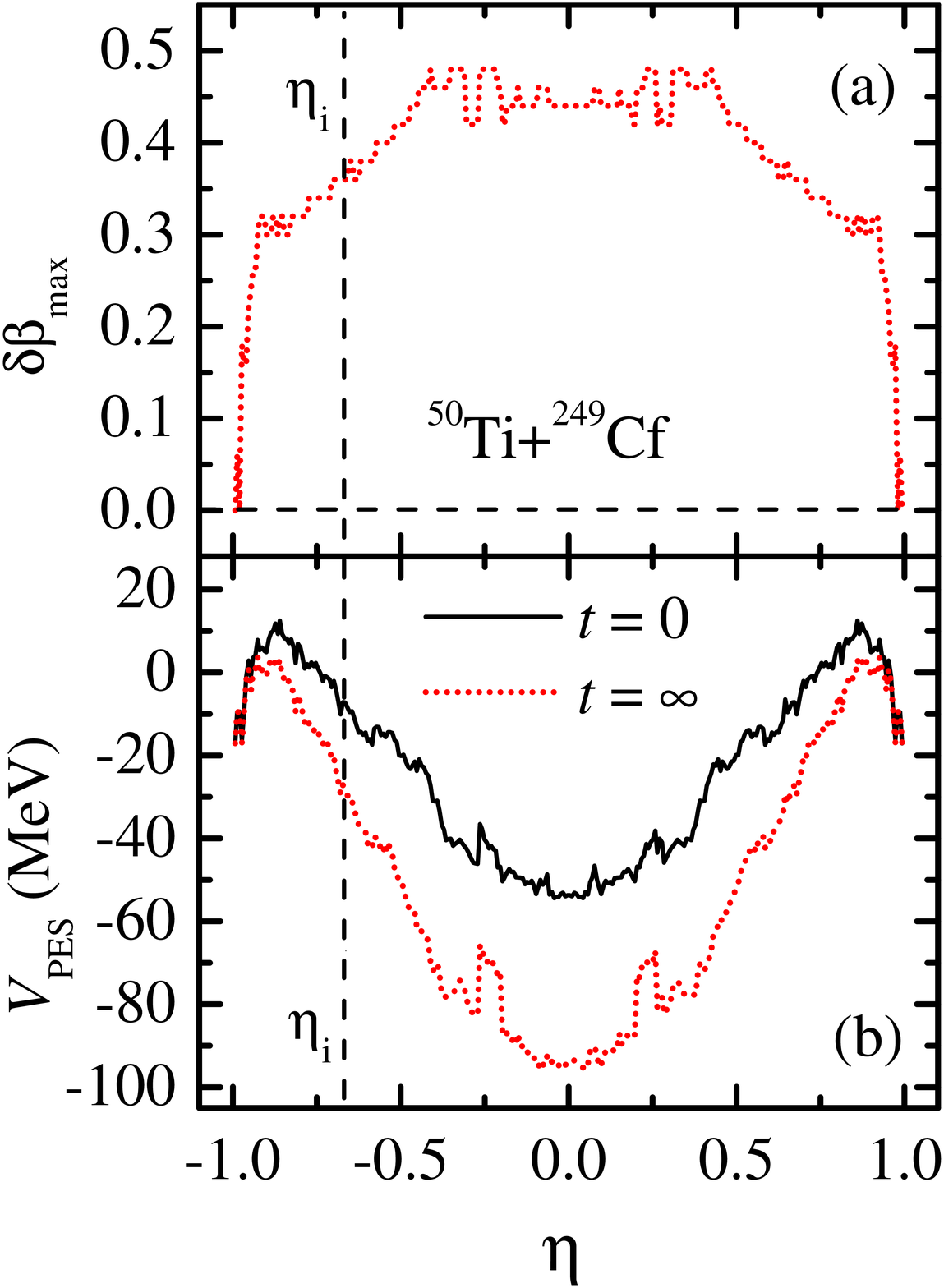}
\par\end{centering}
\caption{\label{fig:DyPES}(Color online)
(a) The maximal dynamical deformation $\delta\beta_\mathrm{max}$ and 
(b) the dynamical potential energy surface (DyPES) defined in Eq.~(\ref{eq:PES}) 
at $t=0$ and $t=\infty$ 
as functions of the mass asymmetry coordinate $\eta$ for 
the projectile-target combination $^{50}$Ti + $^{249}$Cf.
The vertical dashed line shows the entrance channel.
}
\end{figure}

Figure~\ref{fig:DyPES} shows
the maximal dynamical deformation $\delta\beta_\mathrm{max}$ and the DyPES
defined in Eq.~(\ref{eq:PES}) at $t=0$ and $t=\infty$ as functions of the mass asymmetry
coordinate $\eta \equiv (A_1 - A_2)/(A_1 + A_2)$ for $^{50}$Ti + $^{249}$Cf.
The values of nuclear masses are taken from Refs.~\cite{Audi2003_NPA729-337, 
Moeller1995_ADNDT59-185}.
In general, with the mass asymmetry increasing, the maximal dynamical deformation 
becomes smaller and when $\eta$ approaches to $\pm 1$, which corresponds to 
the formation of a CN, $\delta\beta_\mathrm{max}$ approaches to zero.
The local excitation energy becomes larger when the dynamical deformation develops.
For those DNS's with larger values of $\delta\beta_\mathrm{max}$, the gain of local
excitation energy, $\delta V^\mathrm{max}_\mathrm{PES}(A_1) \equiv
V_\mathrm{PES}(A_1, t=0) - V_\mathrm{PES}(A_1, t=\infty)$,
is also larger.
Note that, since $\delta\beta_\mathrm{max}$ for $\eta = \pm 1$ is zero, the local
excitation energy of the CN $E^*_\mathrm{CN}$ is always fixed and 
$E^*_\mathrm{CN} = Q + E_\mathrm{c.m.}$ with $Q$ the reaction energy.
In Fig.~\ref{fig:EV-db} 
we plot the potential energy $V_\mathrm{PES}$ (the red dotted curve) as a function of 
$\delta\beta$ for $^{50}$Ti + $^{249}$Cf. 
It can be seen that the potential energy decreases almost linearly with 
the dynamical deformation increasing.
Therefore, in order to reduce the numerical time, we assume
\begin{equation}
 V_\mathrm{PES}(A_1, t)
 =
 V_\mathrm{PES}(A_1, t=0)
 - \frac{\delta\beta(t)}{\delta\beta_\mathrm{max}}
   \delta V^\mathrm{max}_\mathrm{PES}(A_1)
 .
\end{equation}
Note that a dynamical PES has been calculated microscopically to describe a continuous transition 
from the initial diabatic potential to the asymptotic adiabatic one due to a
residual two-body collision~\cite{Diaz-Torres2004_PRC69-021603, Diaz-Torres2006_PRC74-064601}. 
It would be an interesting topic to explore the connection between the DyPES in the present 
work and that proposed in Refs.~\cite{Diaz-Torres2004_PRC69-021603, Diaz-Torres2006_PRC74-064601}.

\begin{figure}
\begin{centering}
\includegraphics[width=0.75\columnwidth]{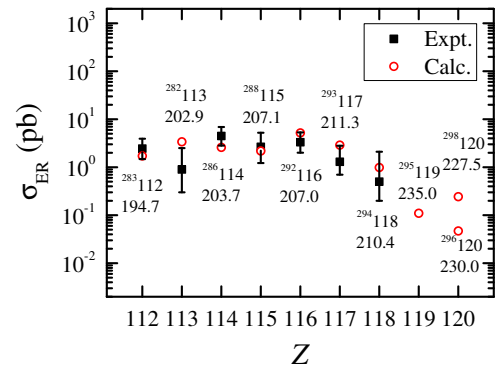}
\par\end{centering}
\caption{\label{fig:112-120}(Color online)
Maximal values of experimentally available evaporation residue cross sections 
for $^{48}$Ca induced reactions leading to SHN with $Z$~=~112--118 compared 
with the theoretical values calculated at the incident energy in the center 
of mass frame $E_\mathrm{c.m.}$ (in MeV) which are indicated in the plot.
The calculated maximal evaporation residue cross sections for $^{50}$Ti 
induced reactions leading to SHN with $Z$ = 119 and 120 are also shown.
For each superheavy nucleus, the charge and mass numbers are given and
the experimental values are shown by black solid squares with error
bars and the theoretical ones by red open circles.
}
\end{figure}

With the DNS-DyPES model, we studied systematically hot fusion reactions
producing superheavy nuclei with the charge number $Z$~=~112--120.
In Fig.~\ref{fig:112-120}, 
maximal values of experimentally available evaporation residue cross sections 
for $^{48}$Ca induced reactions leading to SHN with $Z$~=~112--118~\cite{NRVmisc, 
Oganessian2007_PRC76-011601R, Oganessian2010_PRL104-142502, Oganessian2006_PRC74-044602} 
are compared with the theoretical values calculated at the incident energy in the center 
of mass frame $E_\mathrm{c.m.}$ which are indicated in the plot.
The calculated maximal evaporation residue cross sections for $^{50}$Ti 
induced reactions leading to SHN with $Z$ = 119 and 120 are also shown.
For reactions leading to SHN with $Z$~=~112--118, the projectile is $^{48}$Ca
and targets are $^{238}$U, $^{237}$Np, $^{242}$Pu, $^{243}$Am, $^{248}$Cm,
$^{249}$Bk, and $^{249}$Cf, respectively.
For SHN with $Z$~=~113, 115, and 118, the maximal $\sigma_\mathrm{ER}$
is found in the 3n evaporation residue channel and for those with $Z$~=~114, 116, and 117,
the 4n channel is more favorable.
For the element 112, according to our calculation, the ER cross section in the 4n channel 
leading to $^{282}$Cn is a bit larger than that in the 3n channel.
But the maximal $\sigma_\mathrm{ER}$ is found in the 3n channel in the experiment
and in Fig.~\ref{fig:112-120} the calculated and experimental results for $^{283}$Cn are given. 
From Fig.~\ref{fig:112-120}, one finds a good agreement between the calculation
and the experiment.
We note that the inclusion of the DyPES in this work reduces by about an order 
of magnitude the fusion probability and the ER cross section.

From $Z$~=~112 to 116, the experimental cross section staggers and
the values are between 1 to 10~pb. For $Z>116$, the ER cross
section decreases almost exponentially with the charge number increasing. 
A similar trend is also found in the calculated results for SHN with $Z$~=~112--118.
Whether this trend continues or not is a very interesting question.
The maximal ER cross sections for the SHN $^{296}$119 in $^{50}$Ti +
$^{249}$Bk, $^{296}$120 in $^{50}$Ti + $^{249}$Cf
(the lower circle), and $^{298}$120 in $^{50}$Ti + $^{251}$Cf
(the upper circle) are also shown in Fig.~\ref{fig:112-120}. It
can be seen that, for the element 119, the decreasing tendency in
the ER cross section continues after $Z$~=~118 and the ER cross
section is about 0.11~pb. This tendency continues also for the
element 120 if $^{249}$Cf is used as the target and the ER cross
section is only about 0.05~pb. However, for the 3n-emission channel of 
$^{50}$Ti + $^{251}$Cf, which produces $^{298}$120,
the maximal ER cross section is about 0.25~pb.

\begin{figure}
\begin{centering}
\includegraphics[width=0.70\columnwidth]{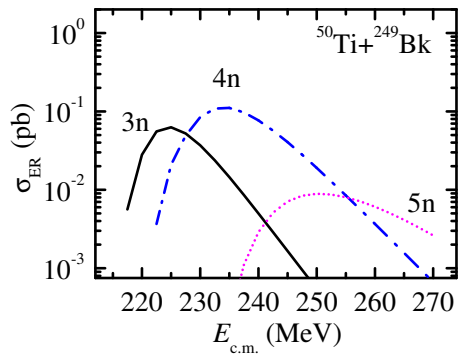}
\par\end{centering}
\caption{\label{fig:Ti50-Bk249}(Color online)
Evaporation residue cross sections $\sigma_\mathrm{ER}$ as a function of the incident energy
in the center of mass frame $E_\mathrm{c.m.}$ for the reaction $^{50}$Ti+$^{249}$Bk.
}
\end{figure}

The synthesis of isotopes of the element 119 with $^{50}$Ti as the projectile 
has been investigated extensively from the theoretical side. 
As we mentioned earlier, the maximal $\sigma_\mathrm{ER}$ from
different models or with different parameters vary very much. 
Let's take the projectile-target combination $^{50}$Ti+$^{249}$Bk as an example. 
The maximal $\sigma_\mathrm{ER}$ varies from 35~\cite{Wang2011_PRC84-061601R}, 
50~\cite{Zagrebaev2008_PRC78-034610}, 
and up to 570 fb~\cite{Liu2011_PRC84-031602R}. 
In this work, the production cross sections for the synthesis of 
the element 119 are studied with the DNS-DyPES model. 
The excitation functions for $^{50}$Ti + $^{249}$Bk leading to $^{294-296}$119 are
represented by solid, dash dot and dotted curves in Fig.~\ref{fig:Ti50-Bk249}. 
The maximal cross sections for 3n and 4n channels are found to be about 
0.06 and 0.11 pb, respectively.

In an attempt to synthesize SHN with $Z$~=~120 using the projectile-target 
combination $^{58}$Fe + $^{244}$Pu, no decay chains consistent with fusion-evaporation
reaction products were observed~\cite{Oganessian2009_PRC79-024603}.
According to the sensitivity of this experiment, the null result sets
an upper limit of 0.4~pb for $\sigma_\mathrm{ER}$.
Some predictions have been made for the ER cross section for this reaction.
For example, the maximal $\sigma_\mathrm{ER}$ is predicted to be about 0.1~pb
in the 3n channel in Ref.~\cite{Nasirov2009_PRC79-024606} and about 0.003~pb
in the 4n channel in Ref.~\cite{Wang2011_PRC84-061601R}.
In this work, the ER cross section in the 4n channel is larger and
the maximal value is only about 4~fb which is far below the current 
experimental limit.
By examining the excitation function for a more asymmetric projectile-target
combination, $^{54}$Cr + $^{248}$Cm, we find that the maximal cross
section is also very small and is only about 6~fb in the 4n channel
which is similar to the results of Ref.~\cite{Wang2011_PRC84-061601R}.

\begin{figure}
\begin{centering}
\includegraphics[width=0.70\columnwidth]{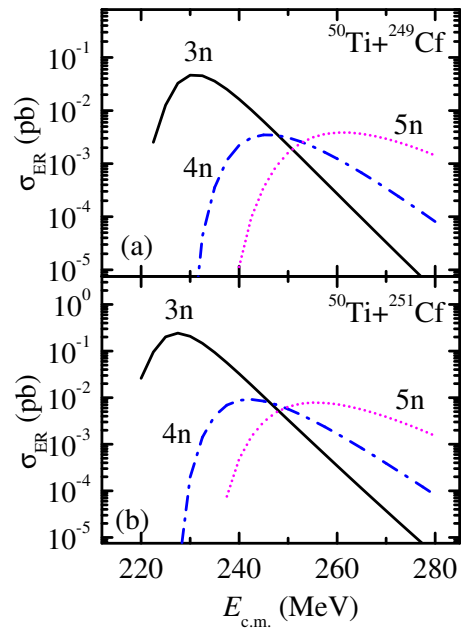}
\par\end{centering}
\caption{\label{fig:Ti50-Cf249-251}(Color online) Evaporation
residue cross sections $\sigma_\mathrm{ER}$ as a function of the
incident energy in the center of mass frame $E_\mathrm{c.m.}$ for 
(a) $^{50}$Ti+$^{249}$Cf (b) and $^{50}$Ti+$^{251}$Cf. }
\end{figure}

Next we investigate $^{50}$Ti induced reactions for the synthesis of
isotopes of the SHE with $Z$~=~120.
The excitation functions for $^{50}$Ti + $^{249}$Cf and $^{50}$Ti + $^{251}$Cf
are shown in Fig.~\ref{fig:Ti50-Cf249-251}.
In both cases the 3n-emission channel gives larger ER cross sections 
than does the 4n channel due to the odd-even effects in the survival probability.
It is found that the maximal cross section for $^{50}$Ti + $^{249}$Cf is about 0.05 pb.
However, for $^{50}$Ti + $^{251}$Cf, the ER cross section can be as large as
0.25 pb which is close to the current experiment limit.
The reason for a large ER cross section in these two reactions is that
the fusion probability increases considerably with the mass asymmetry increasing;
the fusion probabilities for $^{50}$Ti + $^{249,251}$Cf are about one order 
of magnitude larger than that for $^{58}$Fe + $^{244}$Pu~\cite{Wang2012_in-prep}.

In summary, we developed a di-nuclear system (DNS) model with a dynamical
potential energy surface (DyPES)---the DNS-DyPES model. 
In this model, the development of dynamical deformations of the two nuclei 
in a DNS is approximately taken into account. 
This is crucial for the determination of the local excitation energy of 
DNS's involved in the fusion process. 
With the DNS-DyPES model, heavy ion fusion reactions with trans-uranium nuclei
as targets are investigated. 
The calculated evaporation residue (ER) cross sections are in good agreement
with available experimental values for the reactions producing
superheavy nuclei with $Z$~=112--118. 
The projectile-target combination $^{50}$Ti+$^{249}$Bk for synthesizing 
the element 119 is studied and the maximal ER cross section is found
to be about 0.11~pb for the 4n-emission channel.
For projectile-target combinations which lead to the synthesis of SHN with $Z=120$, 
the ER cross section increases with the
mass asymmetry of the entrance channel increasing. 
The ER cross sections for $^{58}$Fe+$^{244}$Pu and $^{54}$Cr+$^{248}$Cm are 
relatively small (less than 10~fb) and those for $^{50}$Ti+$^{249}$Cf and
$^{50}$Ti+$^{251}$Cf are about 0.05 and 0.25~pb, respectively. 


We thank Professor Jun-Qing Li for helpful discussions.
The work was supported by NSF China (10975100, 10979066, 11175252, and 11120101005),
MOST of China (2007CB815000),
the Knowledge Innovation Project of CAS (KJCX2-EW-N01 and KJCX2-YW-N32),
and Deutsche Forschungsgemeinschaft (DFG).
N.W., E.G.Z., and S.G.Z. would like to express their gratitude 
to W. Scheid for the kind hospitality extended to them at Giessen University
where part of this work was done under the support of NSF China and DFG.
Part of the numerical results is obtained on the ScGrid of Supercomputing Center,
CNIC of CAS.


\begin{thebibliography}{48}%
\makeatletter
\providecommand \@ifxundefined [1]{%
 \@ifx{#1\undefined}
}%
\providecommand \@ifnum [1]{%
 \ifnum #1\expandafter \@firstoftwo
 \else \expandafter \@secondoftwo
 \fi
}%
\providecommand \@ifx [1]{%
 \ifx #1\expandafter \@firstoftwo
 \else \expandafter \@secondoftwo
 \fi
}%
\providecommand \natexlab [1]{#1}%
\providecommand \enquote  [1]{``#1''}%
\providecommand \bibnamefont  [1]{#1}%
\providecommand \bibfnamefont [1]{#1}%
\providecommand \citenamefont [1]{#1}%
\providecommand \href@noop [0]{\@secondoftwo}%
\providecommand \href [0]{\begingroup \@sanitize@url \@href}%
\providecommand \@href[1]{\@@startlink{#1}\@@href}%
\providecommand \@@href[1]{\endgroup#1\@@endlink}%
\providecommand \@sanitize@url [0]{\catcode `\\12\catcode `\$12\catcode
  `\&12\catcode `\#12\catcode `\^12\catcode `\_12\catcode `\%12\relax}%
\providecommand \@@startlink[1]{}%
\providecommand \@@endlink[0]{}%
\providecommand \url  [0]{\begingroup\@sanitize@url \@url }%
\providecommand \@url [1]{\endgroup\@href {#1}{\urlprefix }}%
\providecommand \urlprefix  [0]{URL }%
\providecommand \Eprint [0]{\href }%
\providecommand \doibase [0]{http://dx.doi.org/}%
\providecommand \selectlanguage [0]{\@gobble}%
\providecommand \bibinfo  [0]{\@secondoftwo}%
\providecommand \bibfield  [0]{\@secondoftwo}%
\providecommand \translation [1]{[#1]}%
\providecommand \BibitemOpen [0]{}%
\providecommand \bibitemStop [0]{}%
\providecommand \bibitemNoStop [0]{.\EOS\space}%
\providecommand \EOS [0]{\spacefactor3000\relax}%
\providecommand \BibitemShut  [1]{\csname bibitem#1\endcsname}%
\let\auto@bib@innerbib\@empty
\bibitem [{\citenamefont {Hofmann}\ and\ \citenamefont
  {M\"unzenberg}(2000)}]{Hofmann2000_RMP72-733}%
  \BibitemOpen
  \bibfield  {author} {\bibinfo {author} {\bibfnamefont {S.}~\bibnamefont
  {Hofmann}}\ and\ \bibinfo {author} {\bibfnamefont {G.}~\bibnamefont
  {M\"unzenberg}},\ }\href {\doibase 10.1103/RevModPhys.72.733} {\bibfield
  {journal} {\bibinfo  {journal} {Rev. Mod. Phys.}\ }\textbf {\bibinfo {volume}
  {72}},\ \bibinfo {pages} {733} (\bibinfo {year} {2000})}\BibitemShut
  {NoStop}%
\bibitem [{\citenamefont {Morita}\ \emph {et~al.}(2004)\citenamefont {Morita},
  \citenamefont {Morimoto}, \citenamefont {Kaji}, \citenamefont {Akiyama},
  \citenamefont {Goto}, \citenamefont {Haba}, \citenamefont {Ideguchi},
  \citenamefont {Kanungo}, \citenamefont {Katori}, \citenamefont {Koura},
  \citenamefont {Kudo}, \citenamefont {Ohnishi}, \citenamefont {Ozawa},
  \citenamefont {Suda}, \citenamefont {Sueki}, \citenamefont {Xu},
  \citenamefont {Yamaguchi}, \citenamefont {Yoneda}, \citenamefont {Yoshida},\
  and\ \citenamefont {Zhao}}]{Morita2004_JPSJ73-2593}%
  \BibitemOpen
  \bibfield  {author} {\bibinfo {author} {\bibfnamefont {K.}~\bibnamefont
  {Morita}}, \bibinfo {author} {\bibfnamefont {K.}~\bibnamefont {Morimoto}},
  \bibinfo {author} {\bibfnamefont {D.}~\bibnamefont {Kaji}}, \bibinfo {author}
  {\bibfnamefont {T.}~\bibnamefont {Akiyama}}, \bibinfo {author} {\bibfnamefont
  {S.-i.}\ \bibnamefont {Goto}}, \bibinfo {author} {\bibfnamefont
  {H.}~\bibnamefont {Haba}}, \bibinfo {author} {\bibfnamefont {E.}~\bibnamefont
  {Ideguchi}}, \bibinfo {author} {\bibfnamefont {R.}~\bibnamefont {Kanungo}},
  \bibinfo {author} {\bibfnamefont {K.}~\bibnamefont {Katori}}, \bibinfo
  {author} {\bibfnamefont {H.}~\bibnamefont {Koura}}, \bibinfo {author}
  {\bibfnamefont {H.}~\bibnamefont {Kudo}}, \bibinfo {author} {\bibfnamefont
  {T.}~\bibnamefont {Ohnishi}}, \bibinfo {author} {\bibfnamefont
  {A.}~\bibnamefont {Ozawa}}, \bibinfo {author} {\bibfnamefont
  {T.}~\bibnamefont {Suda}}, \bibinfo {author} {\bibfnamefont {K.}~\bibnamefont
  {Sueki}}, \bibinfo {author} {\bibfnamefont {H.-S.}\ \bibnamefont {Xu}},
  \bibinfo {author} {\bibfnamefont {T.}~\bibnamefont {Yamaguchi}}, \bibinfo
  {author} {\bibfnamefont {A.}~\bibnamefont {Yoneda}}, \bibinfo {author}
  {\bibfnamefont {A.}~\bibnamefont {Yoshida}}, \ and\ \bibinfo {author}
  {\bibfnamefont {Y.-L.}\ \bibnamefont {Zhao}},\ }\href {\doibase
  10.1143/JPSJ.73.2593} {\bibfield  {journal} {\bibinfo  {journal} {J. Phys.
  Soc. Jpn.}\ }\textbf {\bibinfo {volume} {73}},\ \bibinfo {pages} {2593}
  (\bibinfo {year} {2004})}\BibitemShut {NoStop}%
\bibitem [{\citenamefont {Oganessian}(2007)}]{Oganessian2007_JPG34-R165}%
  \BibitemOpen
  \bibfield  {author} {\bibinfo {author} {\bibfnamefont {Y.}~\bibnamefont
  {Oganessian}},\ }\href {\doibase 10.1088/0954-3899/34/4/R01} {\bibfield
  {journal} {\bibinfo  {journal} {J. Phys. G: Nucl. Phys.}\ }\textbf {\bibinfo
  {volume} {34}},\ \bibinfo {pages} {R165} (\bibinfo {year}
  {2007})}\BibitemShut {NoStop}%
\bibitem [{\citenamefont {Oganessian}\ \emph {et~al.}(2010)\citenamefont
  {Oganessian}, \citenamefont {Abdullin}, \citenamefont {Bailey}, \citenamefont
  {Benker}, \citenamefont {Bennett}, \citenamefont {Dmitriev}, \citenamefont
  {Ezold}, \citenamefont {Hamilton}, \citenamefont {Henderson}, \citenamefont
  {Itkis}, \citenamefont {Lobanov}, \citenamefont {Mezentsev}, \citenamefont
  {Moody}, \citenamefont {Nelson}, \citenamefont {Polyakov}, \citenamefont
  {Porter}, \citenamefont {Ramayya}, \citenamefont {Riley}, \citenamefont
  {Roberto}, \citenamefont {Ryabinin}, \citenamefont {Rykaczewski},
  \citenamefont {Sagaidak}, \citenamefont {Shaughnessy}, \citenamefont
  {Shirokovsky}, \citenamefont {Stoyer}, \citenamefont {Subbotin},
  \citenamefont {Sudowe}, \citenamefont {Sukhov}, \citenamefont {Tsyganov},
  \citenamefont {Utyonkov}, \citenamefont {Voinov}, \citenamefont {Vostokin},\
  and\ \citenamefont {Wilk}}]{Oganessian2010_PRL104-142502}%
  \BibitemOpen
  \bibfield  {author} {\bibinfo {author} {\bibfnamefont {Y.~T.}\ \bibnamefont
  {Oganessian}}, \bibinfo {author} {\bibfnamefont {F.~S.}\ \bibnamefont
  {Abdullin}}, \bibinfo {author} {\bibfnamefont {P.~D.}\ \bibnamefont
  {Bailey}}, \bibinfo {author} {\bibfnamefont {D.~E.}\ \bibnamefont {Benker}},
  \bibinfo {author} {\bibfnamefont {M.~E.}\ \bibnamefont {Bennett}}, \bibinfo
  {author} {\bibfnamefont {S.~N.}\ \bibnamefont {Dmitriev}}, \bibinfo {author}
  {\bibfnamefont {J.~G.}\ \bibnamefont {Ezold}}, \bibinfo {author}
  {\bibfnamefont {J.~H.}\ \bibnamefont {Hamilton}}, \bibinfo {author}
  {\bibfnamefont {R.~A.}\ \bibnamefont {Henderson}}, \bibinfo {author}
  {\bibfnamefont {M.~G.}\ \bibnamefont {Itkis}}, \bibinfo {author}
  {\bibfnamefont {Y.~V.}\ \bibnamefont {Lobanov}}, \bibinfo {author}
  {\bibfnamefont {A.~N.}\ \bibnamefont {Mezentsev}}, \bibinfo {author}
  {\bibfnamefont {K.~J.}\ \bibnamefont {Moody}}, \bibinfo {author}
  {\bibfnamefont {S.~L.}\ \bibnamefont {Nelson}}, \bibinfo {author}
  {\bibfnamefont {A.~N.}\ \bibnamefont {Polyakov}}, \bibinfo {author}
  {\bibfnamefont {C.~E.}\ \bibnamefont {Porter}}, \bibinfo {author}
  {\bibfnamefont {A.~V.}\ \bibnamefont {Ramayya}}, \bibinfo {author}
  {\bibfnamefont {F.~D.}\ \bibnamefont {Riley}}, \bibinfo {author}
  {\bibfnamefont {J.~B.}\ \bibnamefont {Roberto}}, \bibinfo {author}
  {\bibfnamefont {M.~A.}\ \bibnamefont {Ryabinin}}, \bibinfo {author}
  {\bibfnamefont {K.~P.}\ \bibnamefont {Rykaczewski}}, \bibinfo {author}
  {\bibfnamefont {R.~N.}\ \bibnamefont {Sagaidak}}, \bibinfo {author}
  {\bibfnamefont {D.~A.}\ \bibnamefont {Shaughnessy}}, \bibinfo {author}
  {\bibfnamefont {I.~V.}\ \bibnamefont {Shirokovsky}}, \bibinfo {author}
  {\bibfnamefont {M.~A.}\ \bibnamefont {Stoyer}}, \bibinfo {author}
  {\bibfnamefont {V.~G.}\ \bibnamefont {Subbotin}}, \bibinfo {author}
  {\bibfnamefont {R.}~\bibnamefont {Sudowe}}, \bibinfo {author} {\bibfnamefont
  {A.~M.}\ \bibnamefont {Sukhov}}, \bibinfo {author} {\bibfnamefont {Y.~S.}\
  \bibnamefont {Tsyganov}}, \bibinfo {author} {\bibfnamefont {V.~K.}\
  \bibnamefont {Utyonkov}}, \bibinfo {author} {\bibfnamefont {A.~A.}\
  \bibnamefont {Voinov}}, \bibinfo {author} {\bibfnamefont {G.~K.}\
  \bibnamefont {Vostokin}}, \ and\ \bibinfo {author} {\bibfnamefont {P.~A.}\
  \bibnamefont {Wilk}},\ }\href {\doibase 10.1103/PhysRevLett.104.142502}
  {\bibfield  {journal} {\bibinfo  {journal} {Phys. Rev. Lett.}\ }\textbf
  {\bibinfo {volume} {104}},\ \bibinfo {pages} {142502} (\bibinfo {year}
  {2010})}\BibitemShut {NoStop}%
\bibitem [{\citenamefont {Oganessian}\ \emph {et~al.}(2009)\citenamefont
  {Oganessian}, \citenamefont {Utyonkov}, \citenamefont {Lobanov},
  \citenamefont {Abdullin}, \citenamefont {Polyakov}, \citenamefont {Sagaidak},
  \citenamefont {Shirokovsky}, \citenamefont {Tsyganov}, \citenamefont
  {Voinov}, \citenamefont {Mezentsev}, \citenamefont {Subbotin}, \citenamefont
  {Sukhov}, \citenamefont {Subotic}, \citenamefont {Zagrebaev}, \citenamefont
  {Dmitriev}, \citenamefont {Henderson}, \citenamefont {Moody}, \citenamefont
  {Kenneally}, \citenamefont {Landrum}, \citenamefont {Shaughnessy},
  \citenamefont {Stoyer}, \citenamefont {Stoyer},\ and\ \citenamefont
  {Wilk}}]{Oganessian2009_PRC79-024603}%
  \BibitemOpen
  \bibfield  {author} {\bibinfo {author} {\bibfnamefont {Y.~T.}\ \bibnamefont
  {Oganessian}}, \bibinfo {author} {\bibfnamefont {V.~K.}\ \bibnamefont
  {Utyonkov}}, \bibinfo {author} {\bibfnamefont {Y.~V.}\ \bibnamefont
  {Lobanov}}, \bibinfo {author} {\bibfnamefont {F.~S.}\ \bibnamefont
  {Abdullin}}, \bibinfo {author} {\bibfnamefont {A.~N.}\ \bibnamefont
  {Polyakov}}, \bibinfo {author} {\bibfnamefont {R.~N.}\ \bibnamefont
  {Sagaidak}}, \bibinfo {author} {\bibfnamefont {I.~V.}\ \bibnamefont
  {Shirokovsky}}, \bibinfo {author} {\bibfnamefont {Y.~S.}\ \bibnamefont
  {Tsyganov}}, \bibinfo {author} {\bibfnamefont {A.~A.}\ \bibnamefont
  {Voinov}}, \bibinfo {author} {\bibfnamefont {A.~N.}\ \bibnamefont
  {Mezentsev}}, \bibinfo {author} {\bibfnamefont {V.~G.}\ \bibnamefont
  {Subbotin}}, \bibinfo {author} {\bibfnamefont {A.~M.}\ \bibnamefont
  {Sukhov}}, \bibinfo {author} {\bibfnamefont {K.}~\bibnamefont {Subotic}},
  \bibinfo {author} {\bibfnamefont {V.~I.}\ \bibnamefont {Zagrebaev}}, \bibinfo
  {author} {\bibfnamefont {S.~N.}\ \bibnamefont {Dmitriev}}, \bibinfo {author}
  {\bibfnamefont {R.~A.}\ \bibnamefont {Henderson}}, \bibinfo {author}
  {\bibfnamefont {K.~J.}\ \bibnamefont {Moody}}, \bibinfo {author}
  {\bibfnamefont {J.~M.}\ \bibnamefont {Kenneally}}, \bibinfo {author}
  {\bibfnamefont {J.~H.}\ \bibnamefont {Landrum}}, \bibinfo {author}
  {\bibfnamefont {D.~A.}\ \bibnamefont {Shaughnessy}}, \bibinfo {author}
  {\bibfnamefont {M.~A.}\ \bibnamefont {Stoyer}}, \bibinfo {author}
  {\bibfnamefont {N.~J.}\ \bibnamefont {Stoyer}}, \ and\ \bibinfo {author}
  {\bibfnamefont {P.~A.}\ \bibnamefont {Wilk}},\ }\href {\doibase
  10.1103/PhysRevC.79.024603} {\bibfield  {journal} {\bibinfo  {journal} {Phys.
  Rev. C}\ }\textbf {\bibinfo {volume} {79}},\ \bibinfo {pages} {024603}
  (\bibinfo {year} {2009})}\BibitemShut {NoStop}%
\bibitem [{\citenamefont {Dullmann}()}]{Dullmann2011_TASCA11}%
  \BibitemOpen
  \bibfield  {author} {\bibinfo {author} {\bibfnamefont {C.~E.}\ \bibnamefont
  {Dullmann}},\ }\href {http://www-win.gsi.de/tasca11/} {\enquote {\bibinfo
  {title} {News from {TASCA}},}\ }\bibinfo {note} {{talk} given at the 10th
  Workshop on Recoil Separator for Superheavy Element Chemistry, October 14,
  2011, {GSI} {Darmstadt}, Germany}\BibitemShut {NoStop}%
\bibitem [{\citenamefont {Feng}\ \emph {et~al.}(2007)\citenamefont {Feng},
  \citenamefont {Jin}, \citenamefont {Li},\ and\ \citenamefont
  {Scheid}}]{Feng2007_PRC76-044606}%
  \BibitemOpen
  \bibfield  {author} {\bibinfo {author} {\bibfnamefont {Z.-Q.}\ \bibnamefont
  {Feng}}, \bibinfo {author} {\bibfnamefont {G.-M.}\ \bibnamefont {Jin}},
  \bibinfo {author} {\bibfnamefont {J.-Q.}\ \bibnamefont {Li}}, \ and\ \bibinfo
  {author} {\bibfnamefont {W.}~\bibnamefont {Scheid}},\ }\href {\doibase
  10.1103/PhysRevC.76.044606} {\bibfield  {journal} {\bibinfo  {journal} {Phys.
  Rev. C}\ }\textbf {\bibinfo {volume} {76}},\ \bibinfo {pages} {044606}
  (\bibinfo {year} {2007})}\BibitemShut {NoStop}%
\bibitem [{\citenamefont {Nasirov}\ \emph {et~al.}(2009)\citenamefont
  {Nasirov}, \citenamefont {Giardina}, \citenamefont {Mandaglio}, \citenamefont
  {Manganaro}, \citenamefont {Hanappe}, \citenamefont {Heinz}, \citenamefont
  {Hofmann}, \citenamefont {Muminov},\ and\ \citenamefont
  {Scheid}}]{Nasirov2009_PRC79-024606}%
  \BibitemOpen
  \bibfield  {author} {\bibinfo {author} {\bibfnamefont {A.~K.}\ \bibnamefont
  {Nasirov}}, \bibinfo {author} {\bibfnamefont {G.}~\bibnamefont {Giardina}},
  \bibinfo {author} {\bibfnamefont {G.}~\bibnamefont {Mandaglio}}, \bibinfo
  {author} {\bibfnamefont {M.}~\bibnamefont {Manganaro}}, \bibinfo {author}
  {\bibfnamefont {F.}~\bibnamefont {Hanappe}}, \bibinfo {author} {\bibfnamefont
  {S.}~\bibnamefont {Heinz}}, \bibinfo {author} {\bibfnamefont
  {S.}~\bibnamefont {Hofmann}}, \bibinfo {author} {\bibfnamefont {A.~I.}\
  \bibnamefont {Muminov}}, \ and\ \bibinfo {author} {\bibfnamefont
  {W.}~\bibnamefont {Scheid}},\ }\href {\doibase 10.1103/PhysRevC.79.024606}
  {\bibfield  {journal} {\bibinfo  {journal} {Phys. Rev. C}\ }\textbf {\bibinfo
  {volume} {79}},\ \bibinfo {pages} {024606} (\bibinfo {year}
  {2009})}\BibitemShut {NoStop}%
\bibitem [{\citenamefont {Adamian}\ \emph {et~al.}(2009)\citenamefont
  {Adamian}, \citenamefont {Antonenko},\ and\ \citenamefont
  {Scheid}}]{Adamian2009_EPJA41-235}%
  \BibitemOpen
  \bibfield  {author} {\bibinfo {author} {\bibfnamefont {G.}~\bibnamefont
  {Adamian}}, \bibinfo {author} {\bibfnamefont {N.}~\bibnamefont {Antonenko}},
  \ and\ \bibinfo {author} {\bibfnamefont {W.}~\bibnamefont {Scheid}},\ }\href
  {\doibase 10.1140/epja/i2009-10795-4} {\bibfield  {journal} {\bibinfo
  {journal} {Eur. Phys. J. A}\ }\textbf {\bibinfo {volume} {41}},\ \bibinfo
  {pages} {235} (\bibinfo {year} {2009})}\BibitemShut {NoStop}%
\bibitem [{\citenamefont {Gan}\ \emph {et~al.}(2011)\citenamefont {Gan},
  \citenamefont {Zhou}, \citenamefont {Huang}, \citenamefont {Feng},\ and\
  \citenamefont {Li}}]{Gan2011_SciChinaPMA54S1-61}%
  \BibitemOpen
  \bibfield  {author} {\bibinfo {author} {\bibfnamefont {Z.-G.}\ \bibnamefont
  {Gan}}, \bibinfo {author} {\bibfnamefont {X.-H.}\ \bibnamefont {Zhou}},
  \bibinfo {author} {\bibfnamefont {M.-H.}\ \bibnamefont {Huang}}, \bibinfo
  {author} {\bibfnamefont {Z.-Q.}\ \bibnamefont {Feng}}, \ and\ \bibinfo
  {author} {\bibfnamefont {J.-Q.}\ \bibnamefont {Li}},\ }\href {\doibase
  10.1007/s11433-011-4436-4} {\bibfield  {journal} {\bibinfo  {journal} {Sci.
  China-Phys. Mech. Astron.}\ }\textbf {\bibinfo {volume} {54 (Supp. 1)}},\
  \bibinfo {pages} {s61} (\bibinfo {year} {2011})}\BibitemShut {NoStop}%
\bibitem [{\citenamefont {Nasirov}\ \emph {et~al.}(2011)\citenamefont
  {Nasirov}, \citenamefont {Mandaglio}, \citenamefont {Giardina}, \citenamefont
  {Sobiczewski},\ and\ \citenamefont {Muminov}}]{Nasirov2011_PRC84-044612}%
  \BibitemOpen
  \bibfield  {author} {\bibinfo {author} {\bibfnamefont {A.~K.}\ \bibnamefont
  {Nasirov}}, \bibinfo {author} {\bibfnamefont {G.}~\bibnamefont {Mandaglio}},
  \bibinfo {author} {\bibfnamefont {G.}~\bibnamefont {Giardina}}, \bibinfo
  {author} {\bibfnamefont {A.}~\bibnamefont {Sobiczewski}}, \ and\ \bibinfo
  {author} {\bibfnamefont {A.~I.}\ \bibnamefont {Muminov}},\ }\href {\doibase
  10.1103/PhysRevC.84.044612} {\bibfield  {journal} {\bibinfo  {journal} {Phys.
  Rev. C}\ }\textbf {\bibinfo {volume} {84}},\ \bibinfo {pages} {044612}
  (\bibinfo {year} {2011})}\BibitemShut {NoStop}%
\bibitem [{\citenamefont {Liu}\ and\ \citenamefont
  {Bao}(2009)}]{Liu2009_PRC80-054608}%
  \BibitemOpen
  \bibfield  {author} {\bibinfo {author} {\bibfnamefont {Z.~H.}\ \bibnamefont
  {Liu}}\ and\ \bibinfo {author} {\bibfnamefont {J.-D.}\ \bibnamefont {Bao}},\
  }\href {\doibase 10.1103/PhysRevC.80.054608} {\bibfield  {journal} {\bibinfo
  {journal} {Phys. Rev. C}\ }\textbf {\bibinfo {volume} {80}},\ \bibinfo
  {pages} {054608} (\bibinfo {year} {2009})}\BibitemShut {NoStop}%
\bibitem [{\citenamefont {Siwek-Wilczynska}\ \emph {et~al.}(2010)\citenamefont
  {Siwek-Wilczynska}, \citenamefont {Cap},\ and\ \citenamefont
  {Wilczynski}}]{Siwek-Wilczynska2010_IJMPE19-500}%
  \BibitemOpen
  \bibfield  {author} {\bibinfo {author} {\bibfnamefont {K.}~\bibnamefont
  {Siwek-Wilczynska}}, \bibinfo {author} {\bibfnamefont {T.}~\bibnamefont
  {Cap}}, \ and\ \bibinfo {author} {\bibfnamefont {J.}~\bibnamefont
  {Wilczynski}},\ }\href {\doibase 10.1142/S021830131001490X} {\bibfield
  {journal} {\bibinfo  {journal} {Int. J. Mod. Phys. E}\ }\textbf {\bibinfo
  {volume} {19}},\ \bibinfo {pages} {500} (\bibinfo {year} {2010})}\BibitemShut
  {NoStop}%
\bibitem [{\citenamefont {Liu}\ and\ \citenamefont
  {Bao}(2011{\natexlab{a}})}]{Liu2011_PRC83-044613}%
  \BibitemOpen
  \bibfield  {author} {\bibinfo {author} {\bibfnamefont {Z.-H.}\ \bibnamefont
  {Liu}}\ and\ \bibinfo {author} {\bibfnamefont {J.-D.}\ \bibnamefont {Bao}},\
  }\href {\doibase 10.1103/PhysRevC.83.044613} {\bibfield  {journal} {\bibinfo
  {journal} {Phys. Rev. C}\ }\textbf {\bibinfo {volume} {83}},\ \bibinfo
  {pages} {044613} (\bibinfo {year} {2011}{\natexlab{a}})}\BibitemShut
  {NoStop}%
\bibitem [{\citenamefont {Liu}\ and\ \citenamefont
  {Bao}(2011{\natexlab{b}})}]{Liu2011_PRC84-031602R}%
  \BibitemOpen
  \bibfield  {author} {\bibinfo {author} {\bibfnamefont {Z.-H.}\ \bibnamefont
  {Liu}}\ and\ \bibinfo {author} {\bibfnamefont {J.-D.}\ \bibnamefont {Bao}},\
  }\href {\doibase 10.1103/PhysRevC.84.031602} {\bibfield  {journal} {\bibinfo
  {journal} {Phys. Rev. C}\ }\textbf {\bibinfo {volume} {84}},\ \bibinfo
  {pages} {031602(R)} (\bibinfo {year} {2011}{\natexlab{b}})}\BibitemShut
  {NoStop}%
\bibitem [{\citenamefont {Zagrebaev}\ and\ \citenamefont
  {Greiner}(2008)}]{Zagrebaev2008_PRC78-034610}%
  \BibitemOpen
  \bibfield  {author} {\bibinfo {author} {\bibfnamefont {V.}~\bibnamefont
  {Zagrebaev}}\ and\ \bibinfo {author} {\bibfnamefont {W.}~\bibnamefont
  {Greiner}},\ }\href {\doibase 10.1103/PhysRevC.78.034610} {\bibfield
  {journal} {\bibinfo  {journal} {Phys. Rev. C}\ }\textbf {\bibinfo {volume}
  {78}},\ \bibinfo {pages} {034610} (\bibinfo {year} {2008})}\BibitemShut
  {NoStop}%
\bibitem [{\citenamefont {Wang}\ \emph {et~al.}(2011)\citenamefont {Wang},
  \citenamefont {Tian},\ and\ \citenamefont {Scheid}}]{Wang2011_PRC84-061601R}%
  \BibitemOpen
  \bibfield  {author} {\bibinfo {author} {\bibfnamefont {N.}~\bibnamefont
  {Wang}}, \bibinfo {author} {\bibfnamefont {J.}~\bibnamefont {Tian}}, \ and\
  \bibinfo {author} {\bibfnamefont {W.}~\bibnamefont {Scheid}},\ }\href
  {\doibase 10.1103/PhysRevC.84.061601} {\bibfield  {journal} {\bibinfo
  {journal} {Phys. Rev. C}\ }\textbf {\bibinfo {volume} {84}},\ \bibinfo
  {pages} {061601(R)} (\bibinfo {year} {2011})}\BibitemShut {NoStop}%
\bibitem [{\citenamefont {Zhou}()}]{Zhou20100416_Giessen}%
  \BibitemOpen
  \bibfield  {author} {\bibinfo {author} {\bibfnamefont {S.-G.}\ \bibnamefont
  {Zhou}},\ }\href@noop {} {\enquote {\bibinfo {title} {A dinulcear system
  model for fusion reactions with consistent and effective coupled-channel
  effects},}\ }\bibinfo {note} {{seminar} given at Giessen University, April
  16, 2010}\BibitemShut {NoStop}%
\bibitem [{\citenamefont {Wang~\textit{et al.}}()}]{Wang2012_in-prep}%
  \BibitemOpen
  \bibfield  {author} {\bibinfo {author} {\bibfnamefont {N.}~\bibnamefont
  {Wang~\textit{et al.}}},\ }\href@noop {} {}\bibinfo {note} {{in
  preparation}}\BibitemShut {NoStop}%
\bibitem [{\citenamefont {Riedel}\ \emph {et~al.}(1979)\citenamefont {Riedel},
  \citenamefont {Wolschin},\ and\ \citenamefont
  {Noerenberg}}]{Riedel1979_ZPA290-47}%
  \BibitemOpen
  \bibfield  {author} {\bibinfo {author} {\bibfnamefont {C.}~\bibnamefont
  {Riedel}}, \bibinfo {author} {\bibfnamefont {G.}~\bibnamefont {Wolschin}}, \
  and\ \bibinfo {author} {\bibfnamefont {W.}~\bibnamefont {Noerenberg}},\
  }\href {\doibase 10.1007/BF01408479} {\bibfield  {journal} {\bibinfo
  {journal} {Z. Phys. A}\ }\textbf {\bibinfo {volume} {290}},\ \bibinfo {pages}
  {47} (\bibinfo {year} {1979})}\BibitemShut {NoStop}%
\bibitem [{\citenamefont {Riedel}\ and\ \citenamefont
  {N\"orenberg}(1979)}]{Riedel1979_ZPA290-385}%
  \BibitemOpen
  \bibfield  {author} {\bibinfo {author} {\bibfnamefont {C.}~\bibnamefont
  {Riedel}}\ and\ \bibinfo {author} {\bibfnamefont {W.}~\bibnamefont
  {N\"orenberg}},\ }\href {\doibase 10.1007/BF01408400} {\bibfield  {journal}
  {\bibinfo  {journal} {Z. Phys. A}\ }\textbf {\bibinfo {volume} {290}},\
  \bibinfo {pages} {385} (\bibinfo {year} {1979})}\BibitemShut {NoStop}%
\bibitem [{\citenamefont {Wolschin}(1979)}]{Wolschin1979_PLB88-35}%
  \BibitemOpen
  \bibfield  {author} {\bibinfo {author} {\bibfnamefont {G.}~\bibnamefont
  {Wolschin}},\ }\href {\doibase 10.1016/0370-2693(79)90107-2} {\bibfield
  {journal} {\bibinfo  {journal} {Phys. Lett. B}\ }\textbf {\bibinfo {volume}
  {88}},\ \bibinfo {pages} {35} (\bibinfo {year} {1979})}\BibitemShut {NoStop}%
\bibitem [{\citenamefont {Li}\ \emph {et~al.}(2003)\citenamefont {Li},
  \citenamefont {Wang}, \citenamefont {Li}, \citenamefont {Xu}, \citenamefont
  {Zuo}, \citenamefont {Zhao}, \citenamefont {Li},\ and\ \citenamefont
  {Scheid}}]{Li2003_EPL64-750}%
  \BibitemOpen
  \bibfield  {author} {\bibinfo {author} {\bibfnamefont {W.}~\bibnamefont
  {Li}}, \bibinfo {author} {\bibfnamefont {N.}~\bibnamefont {Wang}}, \bibinfo
  {author} {\bibfnamefont {J.~F.}\ \bibnamefont {Li}}, \bibinfo {author}
  {\bibfnamefont {H.}~\bibnamefont {Xu}}, \bibinfo {author} {\bibfnamefont
  {W.}~\bibnamefont {Zuo}}, \bibinfo {author} {\bibfnamefont {E.}~\bibnamefont
  {Zhao}}, \bibinfo {author} {\bibfnamefont {J.~Q.}\ \bibnamefont {Li}}, \ and\
  \bibinfo {author} {\bibfnamefont {W.}~\bibnamefont {Scheid}},\ }\href
  {\doibase 10.1209/epl/i2003-00622-0} {\bibfield  {journal} {\bibinfo
  {journal} {Europhys. Lett.}\ }\textbf {\bibinfo {volume} {64}},\ \bibinfo
  {pages} {750} (\bibinfo {year} {2003})}\BibitemShut {NoStop}%
\bibitem [{\citenamefont {Li}\ \emph {et~al.}(2006)\citenamefont {Li},
  \citenamefont {Wang}, \citenamefont {Jia}, \citenamefont {Xu}, \citenamefont
  {Zuo}, \citenamefont {Li}, \citenamefont {Zhao}, \citenamefont {Li},\ and\
  \citenamefont {Scheid}}]{Li2006_JPG32-1143}%
  \BibitemOpen
  \bibfield  {author} {\bibinfo {author} {\bibfnamefont {W.}~\bibnamefont
  {Li}}, \bibinfo {author} {\bibfnamefont {N.}~\bibnamefont {Wang}}, \bibinfo
  {author} {\bibfnamefont {F.}~\bibnamefont {Jia}}, \bibinfo {author}
  {\bibfnamefont {H.}~\bibnamefont {Xu}}, \bibinfo {author} {\bibfnamefont
  {W.}~\bibnamefont {Zuo}}, \bibinfo {author} {\bibfnamefont {Q.}~\bibnamefont
  {Li}}, \bibinfo {author} {\bibfnamefont {E.}~\bibnamefont {Zhao}}, \bibinfo
  {author} {\bibfnamefont {J.}~\bibnamefont {Li}}, \ and\ \bibinfo {author}
  {\bibfnamefont {W.}~\bibnamefont {Scheid}},\ }\href {\doibase
  10.1088/0954-3899/32/8/006} {\bibfield  {journal} {\bibinfo  {journal} {J.
  Phys. G: Nucl. Phys.}\ }\textbf {\bibinfo {volume} {32}},\ \bibinfo {pages}
  {1143} (\bibinfo {year} {2006})}\BibitemShut {NoStop}%
\bibitem [{\citenamefont {Wang}\ \emph {et~al.}(2008)\citenamefont {Wang},
  \citenamefont {Li},\ and\ \citenamefont {Zhao}}]{Wang2008_PRC78-054607}%
  \BibitemOpen
  \bibfield  {author} {\bibinfo {author} {\bibfnamefont {N.}~\bibnamefont
  {Wang}}, \bibinfo {author} {\bibfnamefont {J.-Q.}\ \bibnamefont {Li}}, \ and\
  \bibinfo {author} {\bibfnamefont {E.-G.}\ \bibnamefont {Zhao}},\ }\href
  {\doibase 10.1103/PhysRevC.78.054607} {\bibfield  {journal} {\bibinfo
  {journal} {Phys. Rev. C}\ }\textbf {\bibinfo {volume} {78}},\ \bibinfo
  {pages} {054607} (\bibinfo {year} {2008})}\BibitemShut {NoStop}%
\bibitem [{\citenamefont {Huang}\ \emph {et~al.}(2011)\citenamefont {Huang},
  \citenamefont {Zhang}, \citenamefont {Gan}, \citenamefont {Zhou},
  \citenamefont {Li},\ and\ \citenamefont {Scheid}}]{Huang2011_PRC84-064619}%
  \BibitemOpen
  \bibfield  {author} {\bibinfo {author} {\bibfnamefont {M.}~\bibnamefont
  {Huang}}, \bibinfo {author} {\bibfnamefont {Z.}~\bibnamefont {Zhang}},
  \bibinfo {author} {\bibfnamefont {Z.}~\bibnamefont {Gan}}, \bibinfo {author}
  {\bibfnamefont {X.}~\bibnamefont {Zhou}}, \bibinfo {author} {\bibfnamefont
  {J.}~\bibnamefont {Li}}, \ and\ \bibinfo {author} {\bibfnamefont
  {W.}~\bibnamefont {Scheid}},\ }\href {\doibase 10.1103/PhysRevC.84.064619}
  {\bibfield  {journal} {\bibinfo  {journal} {Phys. Rev. C}\ }\textbf {\bibinfo
  {volume} {84}},\ \bibinfo {pages} {064619} (\bibinfo {year}
  {2011})}\BibitemShut {NoStop}%
\bibitem [{\citenamefont {Zagrebaev}(2001)}]{Zagrebaev2001_PRC64-034606}%
  \BibitemOpen
  \bibfield  {author} {\bibinfo {author} {\bibfnamefont {V.~I.}\ \bibnamefont
  {Zagrebaev}},\ }\href {\doibase 10.1103/PhysRevC.64.034606} {\bibfield
  {journal} {\bibinfo  {journal} {Phys. Rev. C}\ }\textbf {\bibinfo {volume}
  {64}},\ \bibinfo {pages} {034606} (\bibinfo {year} {2001})}\BibitemShut
  {NoStop}%
\bibitem [{\citenamefont {Feng}\ \emph {et~al.}(2006)\citenamefont {Feng},
  \citenamefont {Jin}, \citenamefont {Fu},\ and\ \citenamefont
  {Li}}]{Feng2006_NPA771-50}%
  \BibitemOpen
  \bibfield  {author} {\bibinfo {author} {\bibfnamefont {Z.-Q.}\ \bibnamefont
  {Feng}}, \bibinfo {author} {\bibfnamefont {G.-M.}\ \bibnamefont {Jin}},
  \bibinfo {author} {\bibfnamefont {F.}~\bibnamefont {Fu}}, \ and\ \bibinfo
  {author} {\bibfnamefont {J.-Q.}\ \bibnamefont {Li}},\ }\href {\doibase
  10.1016/j.nuclphysa.2006.03.002} {\bibfield  {journal} {\bibinfo  {journal}
  {Nucl. Phys. A}\ }\textbf {\bibinfo {volume} {771}},\ \bibinfo {pages} {50}
  (\bibinfo {year} {2006})}\BibitemShut {NoStop}%
\bibitem [{\citenamefont {Zubov}\ \emph {et~al.}(2002)\citenamefont {Zubov},
  \citenamefont {Adamian}, \citenamefont {Antonenko}, \citenamefont {Ivanova},\
  and\ \citenamefont {Scheid}}]{Zubov2002_PRC65-024308}%
  \BibitemOpen
  \bibfield  {author} {\bibinfo {author} {\bibfnamefont {A.~S.}\ \bibnamefont
  {Zubov}}, \bibinfo {author} {\bibfnamefont {G.~G.}\ \bibnamefont {Adamian}},
  \bibinfo {author} {\bibfnamefont {N.~V.}\ \bibnamefont {Antonenko}}, \bibinfo
  {author} {\bibfnamefont {S.~P.}\ \bibnamefont {Ivanova}}, \ and\ \bibinfo
  {author} {\bibfnamefont {W.}~\bibnamefont {Scheid}},\ }\href {\doibase
  10.1103/PhysRevC.65.024308} {\bibfield  {journal} {\bibinfo  {journal} {Phys.
  Rev. C}\ }\textbf {\bibinfo {volume} {65}},\ \bibinfo {pages} {024308}
  (\bibinfo {year} {2002})}\BibitemShut {NoStop}%
\bibitem [{\citenamefont {Xia}\ \emph {et~al.}(2011)\citenamefont {Xia},
  \citenamefont {Sun}, \citenamefont {Zhao},\ and\ \citenamefont
  {Zhou}}]{Xia2011_SciChinaPAM54S1-109}%
  \BibitemOpen
  \bibfield  {author} {\bibinfo {author} {\bibfnamefont {C.-J.}\ \bibnamefont
  {Xia}}, \bibinfo {author} {\bibfnamefont {B.-X.}\ \bibnamefont {Sun}},
  \bibinfo {author} {\bibfnamefont {E.-G.}\ \bibnamefont {Zhao}}, \ and\
  \bibinfo {author} {\bibfnamefont {S.-G.}\ \bibnamefont {Zhou}},\ }\href
  {\doibase 10.1007/s11433-011-4438-2} {\bibfield  {journal} {\bibinfo
  {journal} {Sci. China-Phys. Mech. Astron.}\ }\textbf {\bibinfo {volume} {54
  (Suppl. 1)}},\ \bibinfo {pages} {109} (\bibinfo {year} {2011})},\ \bibinfo
  {note} {arXiv: 1101.2725 [nucl-th]}\BibitemShut {NoStop}%
\bibitem [{\citenamefont {Volkov}(1978)}]{Volkov1978_PR44-93}%
  \BibitemOpen
  \bibfield  {author} {\bibinfo {author} {\bibfnamefont {V.~V.}\ \bibnamefont
  {Volkov}},\ }\href {\doibase 10.1016/0370-1573(78)90200-4} {\bibfield
  {journal} {\bibinfo  {journal} {Phys. Rep.}\ }\textbf {\bibinfo {volume}
  {44}},\ \bibinfo {pages} {93} (\bibinfo {year} {1978})}\BibitemShut {NoStop}%
\bibitem [{\citenamefont {Antonenko}\ \emph {et~al.}(1993)\citenamefont
  {Antonenko}, \citenamefont {Cherepanov}, \citenamefont {Nasirov},
  \citenamefont {Permjakov},\ and\ \citenamefont
  {Volkov}}]{Antonenko1993_PLB319-425}%
  \BibitemOpen
  \bibfield  {author} {\bibinfo {author} {\bibfnamefont {N.}~\bibnamefont
  {Antonenko}}, \bibinfo {author} {\bibfnamefont {E.}~\bibnamefont
  {Cherepanov}}, \bibinfo {author} {\bibfnamefont {A.}~\bibnamefont {Nasirov}},
  \bibinfo {author} {\bibfnamefont {V.}~\bibnamefont {Permjakov}}, \ and\
  \bibinfo {author} {\bibfnamefont {V.}~\bibnamefont {Volkov}},\ }\href
  {\doibase 10.1016/0370-2693(93)91746-A} {\bibfield  {journal} {\bibinfo
  {journal} {Phys. Lett. B}\ }\textbf {\bibinfo {volume} {319}},\ \bibinfo
  {pages} {425} (\bibinfo {year} {1993})}\BibitemShut {NoStop}%
\bibitem [{\citenamefont {Antonenko}\ \emph {et~al.}(1995)\citenamefont
  {Antonenko}, \citenamefont {Cherepanov}, \citenamefont {Nasirov},
  \citenamefont {Permjakov},\ and\ \citenamefont
  {Volkov}}]{Antonenko1995_PRC51-2635}%
  \BibitemOpen
  \bibfield  {author} {\bibinfo {author} {\bibfnamefont {N.~V.}\ \bibnamefont
  {Antonenko}}, \bibinfo {author} {\bibfnamefont {E.~A.}\ \bibnamefont
  {Cherepanov}}, \bibinfo {author} {\bibfnamefont {A.~K.}\ \bibnamefont
  {Nasirov}}, \bibinfo {author} {\bibfnamefont {V.~P.}\ \bibnamefont
  {Permjakov}}, \ and\ \bibinfo {author} {\bibfnamefont {V.~V.}\ \bibnamefont
  {Volkov}},\ }\href {\doibase 10.1103/PhysRevC.51.2635} {\bibfield  {journal}
  {\bibinfo  {journal} {Phys. Rev. C}\ }\textbf {\bibinfo {volume} {51}},\
  \bibinfo {pages} {2635} (\bibinfo {year} {1995})}\BibitemShut {NoStop}%
\bibitem [{\citenamefont {Adamian}\ \emph {et~al.}(1997)\citenamefont
  {Adamian}, \citenamefont {Antonenko}, \citenamefont {Scheid},\ and\
  \citenamefont {Volkov}}]{Adamian1997_NPA627-361}%
  \BibitemOpen
  \bibfield  {author} {\bibinfo {author} {\bibfnamefont {G.~G.}\ \bibnamefont
  {Adamian}}, \bibinfo {author} {\bibfnamefont {N.~V.}\ \bibnamefont
  {Antonenko}}, \bibinfo {author} {\bibfnamefont {W.}~\bibnamefont {Scheid}}, \
  and\ \bibinfo {author} {\bibfnamefont {V.~V.}\ \bibnamefont {Volkov}},\
  }\href {\doibase 10.1016/S0375-9474(97)00605-2} {\bibfield  {journal}
  {\bibinfo  {journal} {Nucl. Phys. A}\ }\textbf {\bibinfo {volume} {627}},\
  \bibinfo {pages} {361} (\bibinfo {year} {1997})}\BibitemShut {NoStop}%
\bibitem [{\citenamefont {Zhao}\ \emph {et~al.}(2008)\citenamefont {Zhao},
  \citenamefont {Wang}, \citenamefont {Feng}, \citenamefont {Li}, \citenamefont
  {Zhou},\ and\ \citenamefont {Scheid}}]{Zhao2008_IJMPE17-1937}%
  \BibitemOpen
  \bibfield  {author} {\bibinfo {author} {\bibfnamefont {E.~G.}\ \bibnamefont
  {Zhao}}, \bibinfo {author} {\bibfnamefont {N.}~\bibnamefont {Wang}}, \bibinfo
  {author} {\bibfnamefont {Z.~Q.}\ \bibnamefont {Feng}}, \bibinfo {author}
  {\bibfnamefont {J.~Q.}\ \bibnamefont {Li}}, \bibinfo {author} {\bibfnamefont
  {S.~G.}\ \bibnamefont {Zhou}}, \ and\ \bibinfo {author} {\bibfnamefont
  {W.}~\bibnamefont {Scheid}},\ }\href {\doibase 10.1142/S021830130801091X}
  {\bibfield  {journal} {\bibinfo  {journal} {Int. J. Mod. Phys. E}\ }\textbf
  {\bibinfo {volume} {17}},\ \bibinfo {pages} {1937} (\bibinfo {year}
  {2008})}\BibitemShut {NoStop}%
\bibitem [{\citenamefont {Zubov}\ \emph {et~al.}(2009)\citenamefont {Zubov},
  \citenamefont {Adamian},\ and\ \citenamefont
  {Antonenko}}]{Zubov2009_PPN40-847}%
  \BibitemOpen
  \bibfield  {author} {\bibinfo {author} {\bibfnamefont {A.}~\bibnamefont
  {Zubov}}, \bibinfo {author} {\bibfnamefont {G.}~\bibnamefont {Adamian}}, \
  and\ \bibinfo {author} {\bibfnamefont {N.}~\bibnamefont {Antonenko}},\ }\href
  {\doibase 10.1134/S1063779609060057} {\bibfield  {journal} {\bibinfo
  {journal} {Phys. Part. Nucl.}\ }\textbf {\bibinfo {volume} {40}},\ \bibinfo
  {pages} {847} (\bibinfo {year} {2009})}\BibitemShut {NoStop}%
\bibitem [{\citenamefont {Li}\ \emph {et~al.}(2010)\citenamefont {Li},
  \citenamefont {Feng}, \citenamefont {Gan}, \citenamefont {Zhou},
  \citenamefont {Zhang},\ and\ \citenamefont {Scheid}}]{Li2010_NPA834-353c}%
  \BibitemOpen
  \bibfield  {author} {\bibinfo {author} {\bibfnamefont {J.-Q.}\ \bibnamefont
  {Li}}, \bibinfo {author} {\bibfnamefont {Z.-Q.}\ \bibnamefont {Feng}},
  \bibinfo {author} {\bibfnamefont {Z.-G.}\ \bibnamefont {Gan}}, \bibinfo
  {author} {\bibfnamefont {X.-H.}\ \bibnamefont {Zhou}}, \bibinfo {author}
  {\bibfnamefont {H.-F.}\ \bibnamefont {Zhang}}, \ and\ \bibinfo {author}
  {\bibfnamefont {W.}~\bibnamefont {Scheid}},\ }\href {\doibase
  10.1016/j.nuclphysa.2010.01.038} {\bibfield  {journal} {\bibinfo  {journal}
  {Nucl. Phys. A}\ }\textbf {\bibinfo {volume} {834}},\ \bibinfo {pages} {353c}
  (\bibinfo {year} {2010})}\BibitemShut {NoStop}%
\bibitem [{\citenamefont {Feng}\ \emph {et~al.}(2011)\citenamefont {Feng},
  \citenamefont {Jin},\ and\ \citenamefont {Li}}]{Feng2011_NPR28-1}%
  \BibitemOpen
  \bibfield  {author} {\bibinfo {author} {\bibfnamefont {Z.-Q.}\ \bibnamefont
  {Feng}}, \bibinfo {author} {\bibfnamefont {G.-M.}\ \bibnamefont {Jin}}, \
  and\ \bibinfo {author} {\bibfnamefont {J.-Q.}\ \bibnamefont {Li}},\ }\href
  {http://www.npr.ac.cn/qikan/Cpaper/zhaiyao.asp?bsid=16175} {\bibfield
  {journal} {\bibinfo  {journal} {Nucl. Phys. Rev.}\ }\textbf {\bibinfo
  {volume} {28}},\ \bibinfo {pages} {1} (\bibinfo {year} {2011})}\BibitemShut
  {NoStop}%
\bibitem [{\citenamefont {Adamian}\ \emph {et~al.}(1996)\citenamefont
  {Adamian}, \citenamefont {Antonenko}, \citenamefont {Jolos}, \citenamefont
  {Ivanova},\ and\ \citenamefont {Melnikova}}]{Adamian1996_IJMPE5-191}%
  \BibitemOpen
  \bibfield  {author} {\bibinfo {author} {\bibfnamefont {G.~G.}\ \bibnamefont
  {Adamian}}, \bibinfo {author} {\bibfnamefont {N.~V.}\ \bibnamefont
  {Antonenko}}, \bibinfo {author} {\bibfnamefont {R.~V.}\ \bibnamefont
  {Jolos}}, \bibinfo {author} {\bibfnamefont {S.~P.}\ \bibnamefont {Ivanova}},
  \ and\ \bibinfo {author} {\bibfnamefont {O.~I.}\ \bibnamefont {Melnikova}},\
  }\href {\doibase 10.1142/S0218301396000098} {\bibfield  {journal} {\bibinfo
  {journal} {Int. J. Mod. Phys. E}\ }\textbf {\bibinfo {volume} {5}},\ \bibinfo
  {pages} {191} (\bibinfo {year} {1996})}\BibitemShut {NoStop}%
\bibitem [{\citenamefont {Wong}(1973)}]{Wong1973_PRL31-766}%
  \BibitemOpen
  \bibfield  {author} {\bibinfo {author} {\bibfnamefont {C.~Y.}\ \bibnamefont
  {Wong}},\ }\href {\doibase 10.1103/PhysRevLett.31.766} {\bibfield  {journal}
  {\bibinfo  {journal} {Phys. Rev. Lett.}\ }\textbf {\bibinfo {volume} {31}},\
  \bibinfo {pages} {766} (\bibinfo {year} {1973})}\BibitemShut {NoStop}%
\bibitem [{\citenamefont {Myers}\ and\ \citenamefont
  {Swiatecki}(1966)}]{Myers1966_NP81-1}%
  \BibitemOpen
  \bibfield  {author} {\bibinfo {author} {\bibfnamefont {W.~D.}\ \bibnamefont
  {Myers}}\ and\ \bibinfo {author} {\bibfnamefont {W.~J.}\ \bibnamefont
  {Swiatecki}},\ }\href {\doibase 10.1016/0029-5582(66)90639-0} {\bibfield
  {journal} {\bibinfo  {journal} {Nucl. Phys.}\ }\textbf {\bibinfo {volume}
  {81}},\ \bibinfo {pages} {1} (\bibinfo {year} {1966})}\BibitemShut {NoStop}%
\bibitem [{\citenamefont {M\"oller}\ \emph {et~al.}(1995)\citenamefont
  {M\"oller}, \citenamefont {Nix}, \citenamefont {Myers},\ and\ \citenamefont
  {Swiatecki}}]{Moeller1995_ADNDT59-185}%
  \BibitemOpen
  \bibfield  {author} {\bibinfo {author} {\bibfnamefont {P.}~\bibnamefont
  {M\"oller}}, \bibinfo {author} {\bibfnamefont {J.~R.}\ \bibnamefont {Nix}},
  \bibinfo {author} {\bibfnamefont {W.~D.}\ \bibnamefont {Myers}}, \ and\
  \bibinfo {author} {\bibfnamefont {W.~J.}\ \bibnamefont {Swiatecki}},\ }\href
  {\doibase 10.1006/adnd.1995.1002} {\bibfield  {journal} {\bibinfo  {journal}
  {At. Data Nucl. Data Tables}\ }\textbf {\bibinfo {volume} {59}},\ \bibinfo
  {pages} {185} (\bibinfo {year} {1995})}\BibitemShut {NoStop}%
\bibitem [{\citenamefont {Audi}\ \emph {et~al.}(2003)\citenamefont {Audi},
  \citenamefont {Wapstra},\ and\ \citenamefont
  {Thibault}}]{Audi2003_NPA729-337}%
  \BibitemOpen
  \bibfield  {author} {\bibinfo {author} {\bibfnamefont {G.}~\bibnamefont
  {Audi}}, \bibinfo {author} {\bibfnamefont {A.~H.}\ \bibnamefont {Wapstra}}, \
  and\ \bibinfo {author} {\bibfnamefont {C.}~\bibnamefont {Thibault}},\ }\href
  {\doibase 10.1016/j.nuclphysa.2003.11.003} {\bibfield  {journal} {\bibinfo
  {journal} {Nucl. Phys. A}\ }\textbf {\bibinfo {volume} {729}},\ \bibinfo
  {pages} {337} (\bibinfo {year} {2003})}\BibitemShut {NoStop}%
\bibitem [{\citenamefont {Diaz-Torres}(2004)}]{Diaz-Torres2004_PRC69-021603}%
  \BibitemOpen
  \bibfield  {author} {\bibinfo {author} {\bibfnamefont {A.}~\bibnamefont
  {Diaz-Torres}},\ }\href {\doibase 10.1103/PhysRevC.69.021603} {\bibfield
  {journal} {\bibinfo  {journal} {Phys. Rev. C}\ }\textbf {\bibinfo {volume}
  {69}},\ \bibinfo {pages} {021603(R)} (\bibinfo {year} {2004})}\BibitemShut
  {NoStop}%
\bibitem [{\citenamefont {Diaz-Torres}(2006)}]{Diaz-Torres2006_PRC74-064601}%
  \BibitemOpen
  \bibfield  {author} {\bibinfo {author} {\bibfnamefont {A.}~\bibnamefont
  {Diaz-Torres}},\ }\href {\doibase 10.1103/PhysRevC.74.064601} {\bibfield
  {journal} {\bibinfo  {journal} {Phys. Rev. C}\ }\textbf {\bibinfo {volume}
  {74}},\ \bibinfo {pages} {064601} (\bibinfo {year} {2006})}\BibitemShut
  {NoStop}%
\bibitem [{\citenamefont {Zagrebaev}\ \emph {et~al.}()\citenamefont
  {Zagrebaev}, \citenamefont {Denikin}, \citenamefont {Karpov}, \citenamefont
  {Alekseev}, \citenamefont {Samarin}, \citenamefont {Rachkov},\ and\
  \citenamefont {Naumenko}}]{NRVmisc}%
  \BibitemOpen
  \bibfield  {author} {\bibinfo {author} {\bibfnamefont {V.~I.}\ \bibnamefont
  {Zagrebaev}}, \bibinfo {author} {\bibfnamefont {A.~S.}\ \bibnamefont
  {Denikin}}, \bibinfo {author} {\bibfnamefont {A.~V.}\ \bibnamefont {Karpov}},
  \bibinfo {author} {\bibfnamefont {A.~P.}\ \bibnamefont {Alekseev}}, \bibinfo
  {author} {\bibfnamefont {V.~V.}\ \bibnamefont {Samarin}}, \bibinfo {author}
  {\bibfnamefont {V.~A.}\ \bibnamefont {Rachkov}}, \ and\ \bibinfo {author}
  {\bibfnamefont {M.~A.}\ \bibnamefont {Naumenko}},\ }\href
  {http://nrv.jinr.ru/nrv/} {\enquote {\bibinfo {title} {Low-energy nuclear
  knowledge base ({N}uclear {R}eaction {V}ideo)},}\ }\bibinfo {howpublished}
  {\url{http://nrv.jinr.ru/nrv/}}\BibitemShut {NoStop}%
\bibitem [{\citenamefont {Oganessian}\ \emph {et~al.}(2007)\citenamefont
  {Oganessian}, \citenamefont {Utyonkov}, \citenamefont {Lobanov},
  \citenamefont {Abdullin}, \citenamefont {Polyakov}, \citenamefont {Sagaidak},
  \citenamefont {Shirokovsky}, \citenamefont {Tsyganov}, \citenamefont
  {Voinov}, \citenamefont {Gulbekian}, \citenamefont {Bogomolov}, \citenamefont
  {Gikal}, \citenamefont {Mezentsev}, \citenamefont {Subbotin}, \citenamefont
  {Sukhov}, \citenamefont {Subotic}, \citenamefont {Zagrebaev}, \citenamefont
  {Vostokin}, \citenamefont {Itkis}, \citenamefont {Henderson}, \citenamefont
  {Kenneally}, \citenamefont {Landrum}, \citenamefont {Moody}, \citenamefont
  {Shaughnessy}, \citenamefont {Stoyer}, \citenamefont {Stoyer},\ and\
  \citenamefont {Wilk}}]{Oganessian2007_PRC76-011601R}%
  \BibitemOpen
  \bibfield  {author} {\bibinfo {author} {\bibfnamefont {Y.~T.}\ \bibnamefont
  {Oganessian}}, \bibinfo {author} {\bibfnamefont {V.~K.}\ \bibnamefont
  {Utyonkov}}, \bibinfo {author} {\bibfnamefont {Y.~V.}\ \bibnamefont
  {Lobanov}}, \bibinfo {author} {\bibfnamefont {F.~S.}\ \bibnamefont
  {Abdullin}}, \bibinfo {author} {\bibfnamefont {A.~N.}\ \bibnamefont
  {Polyakov}}, \bibinfo {author} {\bibfnamefont {R.~N.}\ \bibnamefont
  {Sagaidak}}, \bibinfo {author} {\bibfnamefont {I.~V.}\ \bibnamefont
  {Shirokovsky}}, \bibinfo {author} {\bibfnamefont {Y.~S.}\ \bibnamefont
  {Tsyganov}}, \bibinfo {author} {\bibfnamefont {A.~A.}\ \bibnamefont
  {Voinov}}, \bibinfo {author} {\bibfnamefont {G.~G.}\ \bibnamefont
  {Gulbekian}}, \bibinfo {author} {\bibfnamefont {S.~L.}\ \bibnamefont
  {Bogomolov}}, \bibinfo {author} {\bibfnamefont {B.~N.}\ \bibnamefont
  {Gikal}}, \bibinfo {author} {\bibfnamefont {A.~N.}\ \bibnamefont
  {Mezentsev}}, \bibinfo {author} {\bibfnamefont {V.~G.}\ \bibnamefont
  {Subbotin}}, \bibinfo {author} {\bibfnamefont {A.~M.}\ \bibnamefont
  {Sukhov}}, \bibinfo {author} {\bibfnamefont {K.}~\bibnamefont {Subotic}},
  \bibinfo {author} {\bibfnamefont {V.~I.}\ \bibnamefont {Zagrebaev}}, \bibinfo
  {author} {\bibfnamefont {G.~K.}\ \bibnamefont {Vostokin}}, \bibinfo {author}
  {\bibfnamefont {M.~G.}\ \bibnamefont {Itkis}}, \bibinfo {author}
  {\bibfnamefont {R.~A.}\ \bibnamefont {Henderson}}, \bibinfo {author}
  {\bibfnamefont {J.~M.}\ \bibnamefont {Kenneally}}, \bibinfo {author}
  {\bibfnamefont {J.~H.}\ \bibnamefont {Landrum}}, \bibinfo {author}
  {\bibfnamefont {K.~J.}\ \bibnamefont {Moody}}, \bibinfo {author}
  {\bibfnamefont {D.~A.}\ \bibnamefont {Shaughnessy}}, \bibinfo {author}
  {\bibfnamefont {M.~A.}\ \bibnamefont {Stoyer}}, \bibinfo {author}
  {\bibfnamefont {N.~J.}\ \bibnamefont {Stoyer}}, \ and\ \bibinfo {author}
  {\bibfnamefont {P.~A.}\ \bibnamefont {Wilk}},\ }\href {\doibase
  10.1103/PhysRevC.76.011601} {\bibfield  {journal} {\bibinfo  {journal} {Phys.
  Rev. C}\ }\textbf {\bibinfo {volume} {76}},\ \bibinfo {pages} {011601(R)}
  (\bibinfo {year} {2007})}\BibitemShut {NoStop}%
\bibitem [{\citenamefont {Oganessian}\ \emph {et~al.}(2006)\citenamefont
  {Oganessian}, \citenamefont {Utyonkov}, \citenamefont {Lobanov},
  \citenamefont {Abdullin}, \citenamefont {Polyakov}, \citenamefont {Sagaidak},
  \citenamefont {Shirokovsky}, \citenamefont {Tsyganov}, \citenamefont
  {Voinov}, \citenamefont {Gulbekian}, \citenamefont {Bogomolov}, \citenamefont
  {Gikal}, \citenamefont {Mezentsev}, \citenamefont {Iliev}, \citenamefont
  {Subbotin}, \citenamefont {Sukhov}, \citenamefont {Subotic}, \citenamefont
  {Zagrebaev}, \citenamefont {Vostokin}, \citenamefont {Itkis}, \citenamefont
  {Moody}, \citenamefont {Patin}, \citenamefont {Shaughnessy}, \citenamefont
  {Stoyer}, \citenamefont {Stoyer}, \citenamefont {Wilk}, \citenamefont
  {Kenneally}, \citenamefont {Landrum}, \citenamefont {Wild},\ and\
  \citenamefont {Lougheed}}]{Oganessian2006_PRC74-044602}%
  \BibitemOpen
  \bibfield  {author} {\bibinfo {author} {\bibfnamefont {Y.~T.}\ \bibnamefont
  {Oganessian}}, \bibinfo {author} {\bibfnamefont {V.~K.}\ \bibnamefont
  {Utyonkov}}, \bibinfo {author} {\bibfnamefont {Y.~V.}\ \bibnamefont
  {Lobanov}}, \bibinfo {author} {\bibfnamefont {F.~S.}\ \bibnamefont
  {Abdullin}}, \bibinfo {author} {\bibfnamefont {A.~N.}\ \bibnamefont
  {Polyakov}}, \bibinfo {author} {\bibfnamefont {R.~N.}\ \bibnamefont
  {Sagaidak}}, \bibinfo {author} {\bibfnamefont {I.~V.}\ \bibnamefont
  {Shirokovsky}}, \bibinfo {author} {\bibfnamefont {Y.~S.}\ \bibnamefont
  {Tsyganov}}, \bibinfo {author} {\bibfnamefont {A.~A.}\ \bibnamefont
  {Voinov}}, \bibinfo {author} {\bibfnamefont {G.~G.}\ \bibnamefont
  {Gulbekian}}, \bibinfo {author} {\bibfnamefont {S.~L.}\ \bibnamefont
  {Bogomolov}}, \bibinfo {author} {\bibfnamefont {B.~N.}\ \bibnamefont
  {Gikal}}, \bibinfo {author} {\bibfnamefont {A.~N.}\ \bibnamefont
  {Mezentsev}}, \bibinfo {author} {\bibfnamefont {S.}~\bibnamefont {Iliev}},
  \bibinfo {author} {\bibfnamefont {V.~G.}\ \bibnamefont {Subbotin}}, \bibinfo
  {author} {\bibfnamefont {A.~M.}\ \bibnamefont {Sukhov}}, \bibinfo {author}
  {\bibfnamefont {K.}~\bibnamefont {Subotic}}, \bibinfo {author} {\bibfnamefont
  {V.~I.}\ \bibnamefont {Zagrebaev}}, \bibinfo {author} {\bibfnamefont {G.~K.}\
  \bibnamefont {Vostokin}}, \bibinfo {author} {\bibfnamefont {M.~G.}\
  \bibnamefont {Itkis}}, \bibinfo {author} {\bibfnamefont {K.~J.}\ \bibnamefont
  {Moody}}, \bibinfo {author} {\bibfnamefont {J.~B.}\ \bibnamefont {Patin}},
  \bibinfo {author} {\bibfnamefont {D.~A.}\ \bibnamefont {Shaughnessy}},
  \bibinfo {author} {\bibfnamefont {M.~A.}\ \bibnamefont {Stoyer}}, \bibinfo
  {author} {\bibfnamefont {N.~J.}\ \bibnamefont {Stoyer}}, \bibinfo {author}
  {\bibfnamefont {P.~A.}\ \bibnamefont {Wilk}}, \bibinfo {author}
  {\bibfnamefont {J.~M.}\ \bibnamefont {Kenneally}}, \bibinfo {author}
  {\bibfnamefont {J.~H.}\ \bibnamefont {Landrum}}, \bibinfo {author}
  {\bibfnamefont {J.~F.}\ \bibnamefont {Wild}}, \ and\ \bibinfo {author}
  {\bibfnamefont {R.~W.}\ \bibnamefont {Lougheed}},\ }\href {\doibase
  10.1103/PhysRevC.74.044602} {\bibfield  {journal} {\bibinfo  {journal} {Phys.
  Rev. C}\ }\textbf {\bibinfo {volume} {74}},\ \bibinfo {pages} {044602}
  (\bibinfo {year} {2006})}\BibitemShut {NoStop}%
\end{thebibliography}

%

\end{document}